\documentclass[usenatbib]{mn2e}

\usepackage{amsmath,amssymb, graphics}
\usepackage{pgf,tikz}
\usepackage{bm}
\usepackage{graphicx}
\usepackage{epstopdf}
\usepackage{float}
\usepackage{mathrsfs}

\newcommand{\bTb}{{\bm T}_{\rm b}}
\newcommand{\omb}{\omega_{\rm b}}
\newcommand{\bomb}{\bar \omega_{\rm b}}

\newcommand{\bl}{\bm {\hat l}}
\newcommand{\blb}{\bm {\hat l}_{\rm b}}
\newcommand{\beb}{\bm {e}_{\rm b}}
\newcommand{\bjb}{\bm {j}_{\rm b}}
\newcommand{\eb}{e_{\rm b}}
\newcommand{\ld}{\bm {\hat l}_{\rm d}}
\newcommand{\blone}{{\bm l}_1}
\newcommand{\rin}{r_\text{in}}
\newcommand{\rout}{r_\text{out}}
\newcommand{\Sgin}{\Sigma_\text{in}}
\newcommand{\Md}{M_\text{d}}
\newcommand{\Ld}{L_\text{d}}
\newcommand{\bLd}{{\bm L}_{\rm d}}
\newcommand{\bLb}{{\bm L}_{\rm b}}
\newcommand{\der}{\text{d}}
\newcommand{\pd}{\partial}
\newcommand{\Om}{\Omega}

\newcommand{\lam}{\lambda}

\newcommand{\Lam}{\Lambda}
\newcommand{\kg}{\kappa}

\newcommand{\Icrit}{I_{\rm crit}}

\newcommand{\bG}{{\bm G}}
\newcommand{\gb}{g_{\rm b}}
\newcommand{\fb}{f_{\rm b}}

\newcommand{\tb}{\tau_{\rm b}}
\newcommand{\Vb}{V_{\rm b}}
\newcommand{\Wbb}{W_{\rm bb}}

\newcommand{\cgb}{\gamma_{\rm b}}
\newcommand{\cga}{\gamma_{\rm a}}

\newcommand{\bcdot}{{\bm \cdot}}
\newcommand{\btimes}{{\bm \times}}

\newcommand{\ag}{\alpha}
\newcommand{\bg}{\beta}

\newcommand{\Sg}{\Sigma}

\newcommand{\cs}{c_{\rm s}}
\newcommand{\Mb}{M_{\rm b}}

\newcommand{\ab}{a_{\rm b}}
\newcommand{\mub}{\mu_{\rm b}}

\newcommand{\be}{\begin{equation}}
\newcommand{\ee}{\end{equation}}

\begin{document}

%DL:
\title[Inclination Evolution of Circumbinary Disks]
{Inclination Evolution of Protoplanetary Disks Around Eccentric Binaries}
%{Circumbinary Disk Dynamics around Eccentric Binaries}

\author[J. J. Zanazzi and Dong Lai]{J. J. Zanazzi$^{1}$\thanks{Email: jjz54@cornell.edu}, and Dong Lai$^{1}$ \\
$^{1}$Cornell Center for Astrophysics, Planetary Science, Department of Astronomy, Cornell University, Ithaca, NY 14853, USA}



\maketitle
\begin{abstract}
%DL:
It is usually thought that viscous torque works to align a
circumbinary disk with the binary's orbital plane. However, recent
numerical simulations suggest that the disk may evolve to a
configuration perpendicular to the binary orbit (``polar alignment") if
the binary is eccentric and the initial disk-binary inclination is sufficiently large. We carry out a theoretical study on the long-term evolution of inclined disks around eccentric
binaries, calculating the disk warp profile and dissipative torque
acting on the disk. For disks with aspect ratio $H/r$ larger than the
viscosity parameter $\alpha$, bending wave propagation effectively
makes the disk precess as a quasi-rigid body, while viscosity acts on
the disk warp and twist to drive secular evolution of the disk-binary
inclination.  We derive a simple analytic criterion (in terms of the binary eccentricity and initial disk orientation) for the disk to evolve toward polar alignment with the eccentric binary.  When the disk has a non-negligible angular momentum compared to the binary, the final ``polar alignment" inclination angle is reduced from $90^\circ$.  For typical
protoplanetary disk parameters, the timescale of the inclination
evolution is shorter than the disk lifetime, suggesting that
highly-inclined disks and planets may exist orbiting eccentric
binaries.
%It is usually assumed that viscous disk torques work to align circumbinary disks with the binary's orbital plane.  However, Martin and Lubow recently used smoothed particle hydrodynamics simulations to show a circumbinary disk may evolve to a configuration perpendicular to the binary's orbital plane (polar alignment), if the binary is sufficiently eccentric.  In this work, we generalize previous dynamical studies of disk warping in circumbinary disks to arbitrary mutual binary-disk inclinations and binary eccentricities, calculating the disk warp profile and dissipative viscous torques acting on the disk.  We provide a simple analytic criterion for when a circumbinary disk evolves to be polar with the eccentric binary.  We also calculate the probability a disk will evolve to be polar with the binary, given the initial binary disk inclination and binary eccentricity.  Due to the ubiquity of order unity binary eccentricities in circumbinary disk systems, this work has important implications on the formation and evolution of circumbinary disks, as well as the inclination distribution of circumbinary planets.
\end{abstract}


%DL: Pls add

\section{Introduction}
\label{sec:Intro}

To date, 11 transiting circumbinary planets have been detected around
9 binary star systems
\citep{Doyle(2011),Kostov(2013),Kostov(2014),Kostov(2016),Orosz(2012a),Orosz(2012b),Schwamb(2013),Welsh(2012),Welsh(2015)}.
All planets detected have orbital planes very well aligned with their
binary orbital planes, with mutual binary-planet inclinations not
exceeding $3^\circ$.
%DL: 
The circumbinary planet detectability is a very sensitive function of
the binary-planet inclination \citep{MartinTriaud(2015),Li(2016)}.  If
the mutual inclination is always small ($\lesssim 5^\circ$),
then the occurance rate of circumbinary planets is comperable to that 
of planets around single stars, but if modest
inclinations ($\gtrsim 5^\circ$) are common, the circumbinary planet
occurance rate may be much larger \citep{Armstrong(2014)}.
For these reasons, it is important to understand if and how a binary aligns with its
circumbinary disk from which these planets form.

%To date, 11 circumbinary planets have been detected around 9 binary star systems \citep{Doyle(2011),Kostov(2013),Kostov(2014),Kostov(2016),Orosz(2012a),Orosz(2012b),Schwamb(2013),Welsh(2012),Welsh(2015)}.  All planets detected have orbital planes very well aligned with their binary orbital planes, with mutual binary-planet inclinations not exceeding $3^\circ$.  If small binary planet inclinations are preffered ($\lesssim 5^\circ$), the occurance rate of circumbinary planets is comperable to the occurance rate of planets around single stars, but if marginal binary-planet inclinations are common ($\gtrsim 5^\circ$), the circumbinary planet occurance rate may greatly exceed the occurance rate of planets around single stars \citep{Armstrong(2014)}.  In addition, the circumbinary planet detectability is a very sensitive function of the binary-planet inclination \citep{MartinTriaud(CircBinDiskDyn2015),Li(2016)}.  For these reasons, it is important to understand if and how a binary aligns with a circumbinary disk from which these planets form.

%DL:
Observations show that most circumbinary disks tend to be aligned with their host
binary orbital planes.  The gas rich circumbinary disks HD 98800 B
\citep{Andrews(2010)}, AK Sco \citep{Czekala(2015)}, DQ Tau
\citep{Czekala(2016)}, and the debris circumbiniary disks $\ag$ CrB
and $\bg$ Tri \citep{Kennedy(2012b)} all have mutual disk-binary
inclinations not exceeding $3^\circ$.  
%DL:
{ However, there are some notable exceptions.  The} circumbinary disk around KH 15D is mildly misaligned with
the binary orbital plane by $\sim
10^\circ$-$20^\circ$\citep{Winn(2004),ChiangMurray-Clay(2004),Capelo(2012)}.  
%but this misalignment is not large enough to challenge the general
%picture of aligned circumbinary disks.  
%DL:
{ Shadows \citep{Marino(2015)} and gas kinematics \citep{Casassus(2015)} of the disks in HD 142527 are consistent with a misalignment of $\sim 70^\circ$ between the outer circumbinary disk and binary orbital plane \citep{Lacour(2016)}.  The disks (circumbinary and two circumstellar) in the binary protostar IRS 43 are misaligned with each other and with the binary \citep{Brinch(2016)}.}
Most intriguingly, the debris disk around the eccentric $(\eb=0.77$)
binary 99 Herculis may be highly inclined: By modeling 
the resolved images from \textit{Hershel}, \cite{Kennedy(2012a)} strongly favor 
a disk orientation where the disk
angular momentum vector is inclined to the binary orbital angular
momentum vector by $90^\circ$ (polar alignment).
\cite{Kennedy(2012a)} also produced a model with a disk-binary
inclination of $30^\circ$ which fits the observations, but this
configuration is unlikely, since differential precession of dust due
to the gravitational influence of the binary would rapidly destroy the
disk.

%DL:
Since star/binary formation takes place in turbulent molecular clouds \citep{McKeeOstriker(2007)},
the gas that falls onto the central protostellar core/binary and
assembles onto the disk at different times may rotate in different
directions (e.g. \citealt{Bate(2003)}, see also \citealt{Bate(2010),Fielding(2015)}).
In this scenario, it is reasonable to expect a newly formed binary to be surrounded
by a highly misaligned circumbinary disk which forms as a result of continued
gas accretion \citep{FoucartLai(2013)}. The observed orientations of circumbinary disks
then depend on the long-term inclination evolution driven by binary-disk interactions.

%Because turbulence in a giant molecular cloud may form a circumbinary disk significantly tilted out of the binary orbital plane (e.g. \citealt{Bate(2003)}, see also \citealt{Bate(2010),Fielding(2015)} for generation of misalignments between stellar spins and circumstellar disk angular momentum vectors from molecular cloud turbulence), a way to quickly align the disk angular momentum vector with the binary orbital angular momentum vector is needed to explain observations of circumbinary disks.

%DL:
\cite{FoucartLai(2013),FoucartLai(2014)} studied the warping and
the dissipative torque driving the inclination evolution of a circumbinary
disk, assuming a circular binary. \cite{FoucartLai(2013)} considered
an infinite disk and included the effect of accretion onto the binary,
while \cite{FoucartLai(2014)} considered a more realistic disk of
finite size and angular momentum, which can precess coherently around
the binary.  It was shown that under typical protoplanetary
conditions, both viscous torque associated with disk warping and
accretion torque tend to damp the mutual disk-binary inclination on
timescale much shorter than the disk lifetime (a few Myr).  By
contrast, a circumstellar disc inside a binary can maintain large
misalignment with respect to the binary orbital plane over its entire
lifetime \citep{LubowOgilvie(2000),FoucartLai(2014)}.  This is consistent with the observations that most
circumbinary disks are nearly coplanar with their host binaries. On
the other hand, the observed circumbinary disk misalignment (such as
in KH 15D and IRS 43) can provide useful constraints on the uncertain aspects of
the disc warp theory, such as non-linear effects \citep{Ogilvie(2006)} and parametric instabilities due to disk warping \citep{Gammie(2000),OgilvieLatter(2013)}

%In \cite{FoucartLai(2013),FoucartLai(2014)}, the warp profile and dissipative viscous torques acting on a circumbinary disk were calculated, assuming a circular binary and nearly coplanar circumbinary disk.  \cite{FoucartLai(2013)} considered a disk with infinite angular momentum, while \cite{FoucartLai(2014)} considered a disk with negligible angular momentum compared to the binary orbital angular momentum.  It was shown that viscous torques from disk warping damp the mutual disk-binary inclination over timescales of order 0.5 Myr.  In addition, \cite{FoucartLai(2013)} considered the effect of accretion torques, which align the binary with the disk over similar timescales.  Because the alignment timescale is much shorter than the typical circumbinary disk lifetime (a few Myr), \cite{FoucartLai(2013),FoucartLai(2014)} argued any primordial disk-binary inclination should rapidly decay due to the influence of viscous (and accretion) torques, explaining the nearly coplanar circumbinary disks observed.

%DL: Pls check the wordings in this paragraph
{ However, several recent numerical studies using smoothed particle hydrodynamics (SPH) suggest that other outcomes may be possible for disks around \textit{eccentric} binaries.  \cite{Aly(2015)} showed that disks around binary black holes (which typically lie in the ``viscous regime" of disk warps, with the viscosity parameter $\ag$ larger than the disk aspect ratio $H/r$; \citealt{PapaloizouPringle(1983),Ogilvie(1999)}; see Sec.~\ref{sec:CircBinDiskDyn}) around eccentric binaries may be driven into polar alignment.  }
\cite{MartinLubow(2017)} found numerically that a circumbinary protoplanetary disk {(typically in the bending wave regime, with $\ag \lesssim H/r$ \citealt{PapaloizouLin(1995),LubowOgilvie(2000)})}, inclined to an
eccentric ($\eb=0.5$) binary by $60^\circ$ will evolve to a polar
configuration.  They suggested that this dynamical outcome arises
%This dynamical outcome was argued to arise 
from the combined influence of
the gravitational torque on the disk from the binary and
viscous torques from disk warping.  They also proposed that 99 Herculis (with $\eb=0.77$) 
followed such an evolution to end in the orbital configuration (polar alignment) 
observed today.
% since the binary 99 Herculis has an eccentricity of 0.77.

In this paper, we provide a theoretical anaysis to the above numerical results.  In particular, we generalize the study of \cite{FoucartLai(2014)} to apply to
circumbinary disks with arbitrary disk-binary inclinations and binary
eccentricities.  We derive the critical condition and calculate the
timescale for the disk to evolve toward polar alignment with the
binary.  In Section~\ref{sec:TestPartDyn}, we review the secular
dynamics of a test particle around an eccentric binary.  In
Section~\ref{sec:CircBinDiskDyn}, we calculate the disk warp profile
and dissipative disk torques acting on the disk, and derive the
requirements for the disk to evolve into polar alignment with the
binary.  { Section~\ref{sec:ArbJ} considers the situation when the circumbinary disk has a non-negligible angular momentum compared to the inner binary.}  In Section~\ref{sec:TorqueBinAcc}, we examine the back-reaction torque from the disk on the binary and the effect of gas accretion.  We discuss our results in Section~\ref{sec:Discuss},
and summarize in Section~\ref{sec:Summary}.

%In this work, we generalize the bending perturbative theory of \citep{FoucartLai(2014)} to apply to circumbinary disks with arbitrary disk-binary inclinations and binary eccentricities.  We show analytically the orientation of the disk angular momentum required to drive the disk to be polar to the eccentric binary.  In Section~\ref{sec:TestPartDyn}, we review the secular dynamics of a test particle around an eccentric binary.  In Section~\ref{sec:CircBinDiskDyn}, we calculate the disk warp profile and dissipative disk torques acting on the disk, and derive the requirements for the disk to evolve into polar alignment with the binary.  In Section~\ref{sec:ProbPolar}, we calculate the probability of polar alignment, given the binary-disk inclination and binary eccentricity.  We discuss our results in Section~\ref{sec:Discuss}, and summarize our results in Section~\ref{sec:Summary}.

\section{Test Particle Dynamics}
\label{sec:TestPartDyn}

%DL:
In preparation for later sections, we review the secular dynamics of a test particle
surrounding an eccentric binary \citep{FaragoLaskar(2010),Li(2014),Naoz(2017)}.
% Most of these results are not new \citep{FaragoLaskar(2010),Li(2014),Naoz(2017)}, 
%but we present them in preperation for later sections.  
Consider a circular test mass
with semimajor axis $r$ and orbital angular momentum unit vector
$\bl$, surrounding an eccentric binary with orbital angular momentum
vector $\blb$, eccentricity vector $\beb$, semimajor axis $\ab$, total
mass $\Mb = M_1 + M_2$ (where $M_1,M_2$ are individual masses), and
reduced mass $\mub = M_1 M_2/\Mb$.  The orbit-averaged torque per unit
mass on the test particle is (e.g. \citealt{Liu(2015),Petrovich(2015)})
\be
\bTb = -r^2 n \omb \left[ (1-\eb^2)(\bl \bcdot \blb) \blb \btimes \bl - 5 (\bl \bcdot \beb) \beb \btimes \bl \right],
\label{eq:bTb}
\ee
where $n \simeq \sqrt{G \Mb/r^3}$ is the test particle orbital frequency (mean-motion), and
\be
\omb = \frac{3 G \mu_{\rm b} \ab^2}{4 r^5 n}
\label{eq:omb}
\ee
characterizes the precession frequency of the test particle around the
binary.  The torque $\bTb$ in Eq.~\eqref{eq:bTb} is evaluated to the lowest order in $\ab/r$.

\begin{figure*}
\includegraphics[scale=0.3]{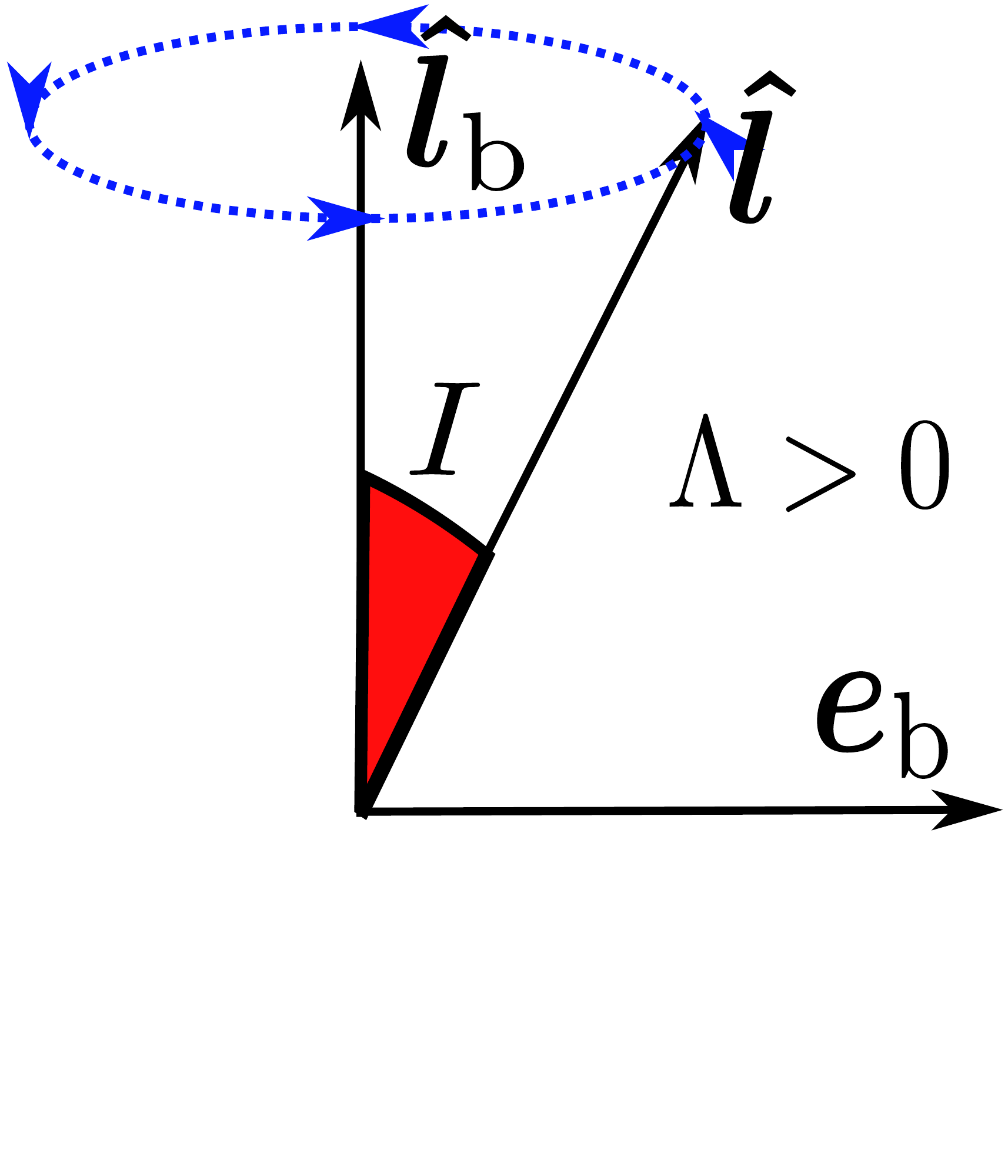}
\hspace{20mm}
\includegraphics[scale=0.3]{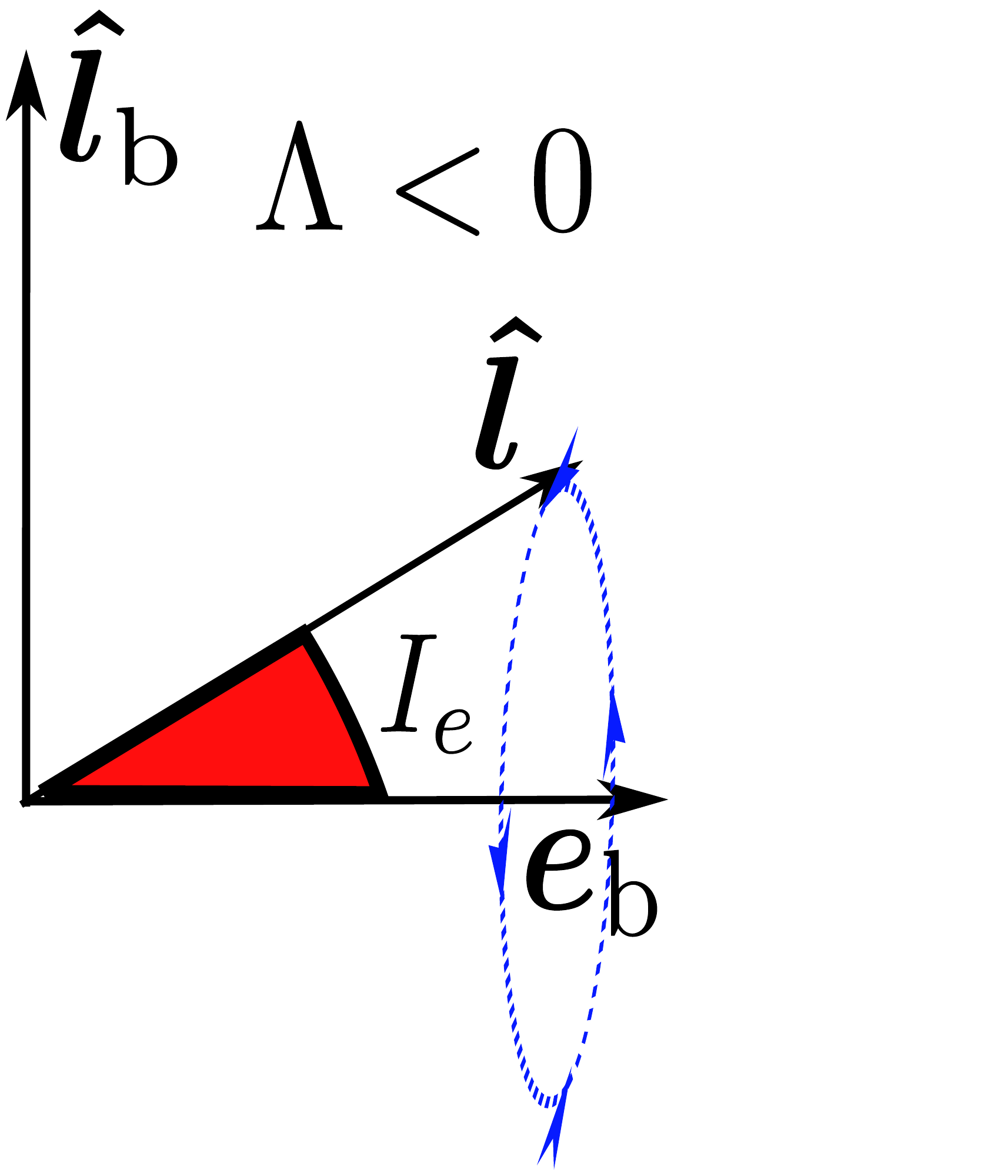}
\caption{Test particle dynamics.  When $\Lam > 0$, $\bl$ precesses 
%DL: typo corrected
around $\blb$, 
%around $\beb$, 
with $I \sim \text{constant}$ and $\Om$
  circulating.  When $\Lam < 0$, $\bl$ precesses around $\beb$, with
  $I_e \sim \text{constant}$ and $\Om_e$ circulating.  See
  Eq.~\eqref{eq:l_coos} for definitions of $I_e$ and $\Om_e$.}
\label{fig:LamDyn}
\end{figure*}

\begin{figure*}
\centering
\includegraphics[scale=1.0]{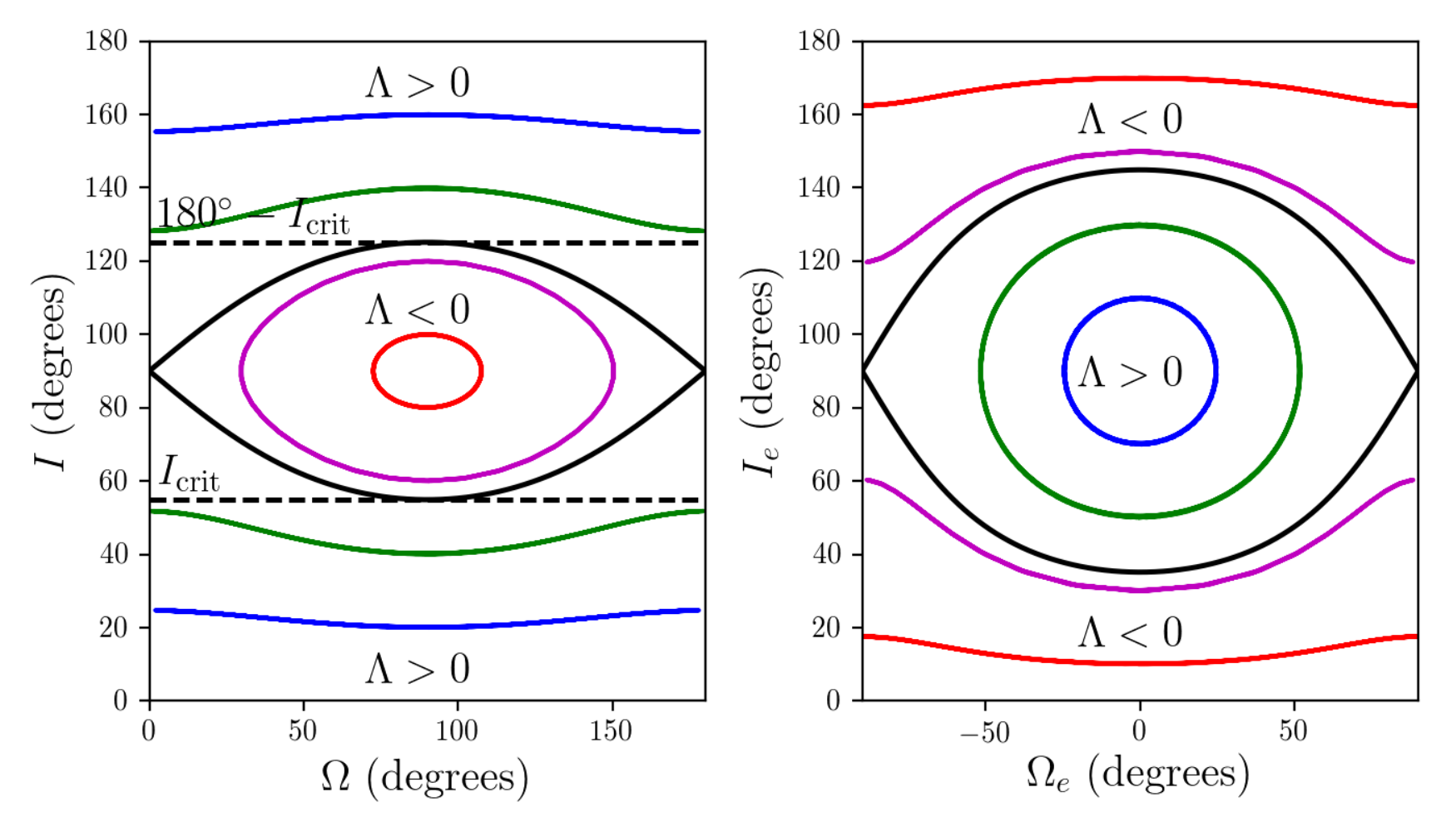}
\caption{ Test particle trajectories in the $I-\Om$ and $I_e-\Om_e$
  planes [see Eq.~\eqref{eq:l_coos}], with binary eccentricity $\eb =0.3$.
  When $\Lam > 0$, $\bl$ precesses around $\blb$, with $I \sim
  \text{constant}$ and $\Om$ circulating its full range of values
  ($0^\circ$-$360^\circ$), while $\Om_e$ librates aroCircBinDiskDynund $0^\circ$.
  When $\Lam < 0$, $\bl$ precesses around $\beb$, with $I_e \sim
  \text{constant}$ and $\Om_e$ circulating its full range of values
  ($-180^\circ$-$180^\circ$), while $\Om$ librating around $90^\circ$
  (see Fig.~\ref{fig:LamDyn}).  Black lines denote the $\Lam = 0$
  separatrix.  The other curves have $\Lam = 0.751$ (blue), $\Lam =
  0.348$ (green), $\Lam = -0.110$ (magenta), $\Lam = -0.409$ (red).
  Only $\Om$ and $\Om_e$ in the range $[0^\circ,180^\circ]$ and
  $[-90^\circ,90^\circ]$ are shown.  The energy curves for $\Om$ in
  $[180^\circ,360^\circ]$ duplicate those of $[0^\circ,180^\circ]$,
  and the energy curves for $\Om_e$ in 
%DL: pls check below
$[90^\circ,270^\circ]$
%$[-180^\circ,90^\circ]$ and  $[90^\circ,180^\circ]$ 
duplicate those of $[-90^\circ,90^\circ]$.}
\label{fig:cons}
\end{figure*}

The time evolution of the test particle's orbital angular momentum vector is given by
\be
\frac{\der \bl}{\der t} = -\omb \left[ (1-\eb^2)(\bl \bcdot \blb) \blb \btimes \bl - 5 (\bl \bcdot \beb) \beb \btimes \bl \right].
\label{eq:dldt_TP}
\ee
Equation~\eqref{eq:dldt_TP} can be solved analytically
\citep{LandauLifshitz(1969),FaragoLaskar(2010),Li(2014)}, but the 
dynamics may be easily understood by analyzing the energy curves.
Equation~\eqref{eq:dldt_TP} has an integral of motion
\be
\Lam = (1-\eb^2)(\bl \bcdot \blb)^2 - 5 (\bl \bcdot \beb)^2,
\label{eq:Lam}
\ee
which is simply related to the quadrupole interaction energy (double-averaged over the two orbits) by
(e.g. \citealt{Tremaine(2009),TremaineYavetz(2014),Liu(2015)})
\be
\Phi_{\rm quad} = \frac{G \mub \ab^2}{8r^3}(1-6 \eb^2 -3\Lambda).
\label{eq:Phi_quad}
\ee
To plot the energy curves, we set up the Cartesian coordinate system $(x,y,z)$, where $\blb = \bm{ \hat z}$ and $\beb = \eb \bm{\hat x}$.  We may write
\begin{align}
\bl &= (\sin I \sin \Om, -\sin I \cos \Om, \cos I)
\nonumber \\
&= (\cos I_e, \sin I_e \sin \Om_e, \sin I_e \cos \Om_e),
\label{eq:l_coos}
\end{align}
where $I$ is the angle between $\bl$ and $\blb$, $\Om$ is the test particle's longitude of the ascending node (measured in the $xy$ plane from the $x$-axis); similarly $I_e$ is the angle between $\bl$ and $\beb$, and $\Om_e$ measures the longitude of the node in the $yz$ plane (see Fig.~\ref{fig:LamDyn}).
%DL: add below
In terms of $I$ and $\Omega$, we have
\begin{equation}
\Lambda=(1-\eb^2)\cos^2I-5\eb^2\sin^2I\sin^2\Omega.
\end{equation}

In Figure~\ref{fig:cons}, we plot the test particle trajectories in
the $I-\Om$ (left panel) and $I_e-\Om_e$ (right panel) planes for
the binary eccentricity $\eb = 0.3$.  The critical separatrix
$\Lam = 0$ is displayed in black in both plots.  When $\Lam > 0$,
$\bl$ precesses around $\blb$ with $I \sim \text{constant}$ and
$\Omega$ circulating the full range ($0-360^\circ$), while $\Omega_e$
librates around $0^\circ$. When $\Lam <0$, $\bl$ precesses around
$\beb$ with $I_e \sim \text{constant}$ and $\Om_e$ circulating the
full range ($0-360^\circ$), while $\Om$ librates around $90^\circ$
(see Fig.~\ref{fig:LamDyn}).

Thus, the test particle angular momentum axis $\bl$ transitions from precession around $\blb$ for $\Lam > 0$ to precession around $\beb$ for $\Lam < 0$.  Because the $\Lam = 0$ separatrix has $\Om \in [0^\circ,360^\circ]$ (Fig.~\ref{fig:cons}), a necessary condition for $\bl$ to precess around $\beb$ is $\Icrit < I < 180^\circ - \Icrit$, where
\be
\Icrit = \cos^{-1} \sqrt{ \frac{5 \eb^2}{1 + 4 \eb^2} }
%DL: add
= \tan^{-1} \sqrt{ \frac{1-\eb^2}{5\eb^2} }.
\label{eq:Icrit}
\ee

Figure~\ref{fig:cons} clearly reveals the stable fixed points of the system. In terms of the variables ($\Omega, I$), the stable fixed points (where $\der I/\der t=\der \Om/\der t=0$) are
$I=\pi/2$ and $\Omega=\pi/2,3\pi/2$, corresponding to $\bl=\pm \beb/\eb$. In terms of the variables ($\Omega_e, I_e$), the fixed points are $I_e=\pi/2$ and $\Omega_e=0,\pi$, corresponding to 
$\bl=\pm \blb$.   We will see in Section 3 that in the presence of dissipation, the disk may be driven toward one of these fixed points.

%%%%%%%%%%%%%%%%%%%%%%%%%%%%%%%%%%%%%%%%%%%%%%%%%%%%%%%%%%
\section{Circumbinary Disk Dynamics}
\label{sec:CircBinDiskDyn}

\begin{figure}
\centering
\includegraphics[scale=0.35]{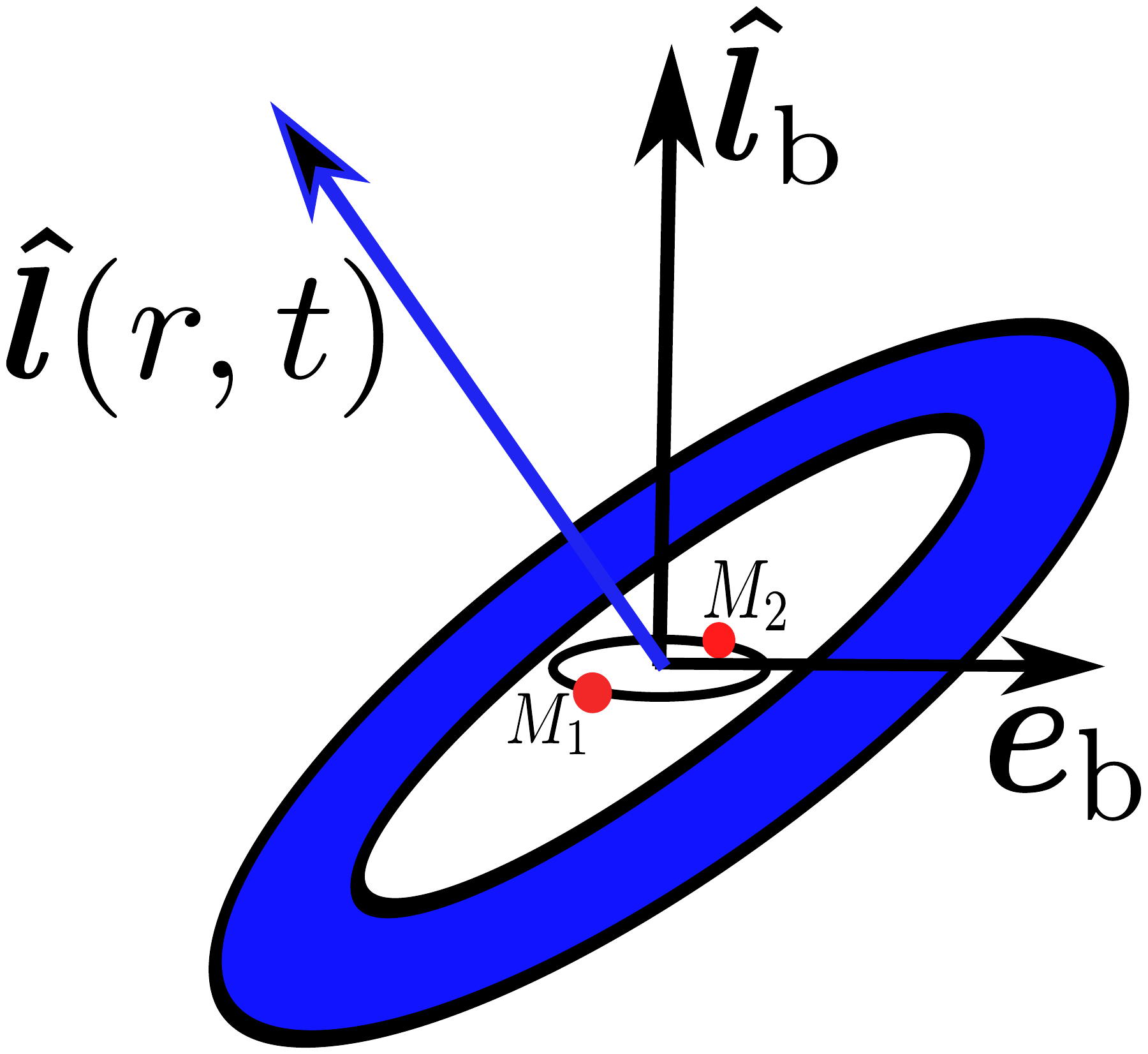}
\caption{Circumbinary disk setup.  The binary has individual masses
  $M_1$ and $M_2$, with total mass $\Mb = M_1 + M_2$ and reduced mass
  $\mub = M_1 M_2/\Mb$, with orbital angular momentum vector $\blb$
  and eccentricity vector $\beb$.  The binary is surrounded by a
  circular circumbinary disk with unit orbital angular momentum $\bl =
  \bl(r,t)$, surface density $\Sg = \Sg(r)$ [Eq.~\eqref{eq:Sg}], and
  inner (outer) truncation radii $\rin$ ($\rout$).}
\label{fig:setup}
\end{figure}

We now consider a binary (with the same parameters as in Section~\ref{sec:TestPartDyn}, see also Fig.~\ref{fig:setup}) surrounded by a circular circumbinary disk with inner truncation radius $\rin$, outer truncation radius $\rout$, with unit angular momentum vector $\bl = \bl(r,t)$, and surface density $\Sg = \Sg(r)$.  For concreteness, we adopt the surface density profile
\be
\Sg(r) = \Sgin \left( \frac{\rin}{r} \right).
\label{eq:Sg}
\ee
We assume $\rin \ll \rout$ throughout this work.  { We could assume a more general surface density profile $\Sg \propto r^{-p}$, with $p$ observationally constrained to lie in the range $0.5-1.5$ (e.g. \citealt{Weidenschilling(1977),WilliamsCieza(2011),ChiangLaughlin(2013)}).  A more general $p$ will effect the disk mass [Eq.~\eqref{eq:Md}] and angular momentum [Eq.~\eqref{eq:Ld}], as well as the precession [Eq.~\eqref{eq:bomb}] and viscous [Eq.~\eqref{eq:cgb}] rates, by factors of order unity.}

The binary has orbital angular momentum
\be
L_{\rm b} = \mub \sqrt{ (1-\eb^2) G \Mb \ab},
\label{eq:Lb}
\ee
while the disk has mass
\begin{align}
\Md &= 2\pi \int_{\rin}^{\rout} \Sg r \der r \simeq 2\pi \Sgin \rin \rout
\label{eq:Md}
\end{align}
and angular momentum (assuming a small disk warp; see below)
\begin{align}
\Ld &= 2\pi \int_{\rin}^{\rout} \Sg r^3 n \der r \simeq \frac{2}{3} \Md \sqrt{G \Mb \rout},
\label{eq:Ld}
\end{align}
where $n(r) \simeq \sqrt{G \Mb/r^3}$.  Comparing $L_{\rm b}$ to $\Ld$, we have
\be
\frac{\Ld}{L_{\rm b}} \simeq 0.067 \, (1-\eb^2)^{-1/2} \left( \frac{\Md}{0.01 \, \mub} \right) \left( \frac{\rout}{100 \, \ab} \right)^{1/2}.
\label{eq:LdOverLb}
\ee
{
Because $L_{\rm b} \gg \Ld$ for typical circumbinary disk parameters,
in this section we assume $\blb$ and $\beb$ are fixed in time, neglecting the
back-reaction torque on the binary from the disk.  We discuss the system's dynamics when $\Ld$ is non-negligible compared to $L_{\rm b}$ in Section~\ref{sec:ArbJ}, and the
effects of the back-reaction torque on the binary from the disk in
Section~\ref{sec:TorqueBinAcc}.
}

\subsection{Qualitative Discussion}
\label{sec:QualDiscuss}

Assuming the disk to be nearly flat, the time evolution of the disk unit angular momentum vector is given by
\be
\frac{\der \bl_{\rm d}}{\der t} = \left \langle \frac{\bTb}{r^2 n} \right \rangle,
\ee
where $\bTb$ is given in Eq.~\eqref{eq:bTb}, $\bl_{\rm d}(t)$ is a suitably averaged unit angular momentum of the disk [see Eq.~\eqref{eq:lexp}], and $\langle \dots \rangle$ implies a proper average over $r$ [see Eq.~\eqref{eq:dl0dt}].  When the disk is flat, the time evolution of $\bl_{\rm d}$ is identical to that of a test particle (see discussion at the end of Sec.~\ref{sec:Form}).

When $\ag \lesssim H/r$ ($H$ is the disk scaleheight, $\ag$ is the viscosity parameter), the main internal torque enforcing disk rigidity and coherent precession comes from bending wave propigation \citep{PapaloizouLin(1995),LubowOgilvie(2000)}.  As bending waves travel at $1/2$ the sound speed, the wave crossing time is of order $t_{\rm bw} = 2 r/\cs$.  When $t_{\rm bw}$ is longer than the characteristic precession time $\omb^{-1}$ [see Eq.~\eqref{eq:omb}], strong disk warps can be induced.  In the extreme nonlinear regime, disk breaking may be possible in circumbinary disks \citep{LarwoodPapaloizou(1997),Facchini(2013),Nixon(2013)}.  To compare $t_{\rm bw}$ with $\omb^{-1}$, we adopt the disk sound speed profile
\be
\cs(r) = H(r) n(r) = h \sqrt{ \frac{G \Mb}{\rin} } \left( \frac{\rin}{r} \right)^{1/2},
\ee
where $h = H/r$.  We find
\be
t_{\rm bw} \omb = 0.94 \left( \frac{0.1}{h} \right) \left( \frac{4 \mub}{\Mb} \right) \left( \frac{2 \, \ab}{\rin} \right)^2 \left( \frac{\rin}{r} \right)^2
\label{eq:tw}
\ee
Thus, we expect that the small warp approximation should be valid everywhere in the disk except the inner-most region.  { Throughout this paper, we scale our results to $h = 0.1$.  Real protoplanetary disks can have aspect ratios in the range $h \sim 0.03-0.2$ (e.g. \citealt{Lynden-BellPringle(1974),ChiangGoldreich(1997),WilliamsCieza(2011)}).  We normalize $\rin$ to $2 \ab$, but note that the inner truncation radius of the disk depends non-trivially on the binary's eccentricity \citep{Miranda(2017)}.}

Although the disk is flat to a good approximation, the interplay between disk twist/warp and viscous dissipation may modify the disk's dynamics over timescales much longer than $\omb^{-1}$. When the external torque $\bTb$ is applied to the disk in the bending wave regime, the disk's viscosity causes the disk to develop a small twist of order
\be
 \frac{\pd \bl_{\rm d}}{\pd \ln r} \bigg|_{\rm twist} \sim - \frac{4 \ag}{\cs^2} \bTb,
\label{eq:dldlnr_visc}
\ee
while the precession of bending waves from a non-Keplarian epicyclic frequency $\kg$ causes the disk to develop a small warp, of order
\be
\frac{\pd \bl_{\rm d}}{\pd \ln r} \bigg|_{\rm warp} \sim - \frac{4}{\cs^2} \left( \frac{\kg^2 - n^2}{2 n^2} \right) \bl_{\rm d} \btimes \bTb.
\label{eq:dldlnr_warp}
\ee

The viscous twist [Eq.~\eqref{eq:dldlnr_visc}] interacts with the external torque, effecting the evolution of $\bl$ over the viscous timescale.  To an order of magnitude, we have
\be
\bigg| \frac{\der \bl_{\rm d}}{\der t} \bigg|_{\rm visc} \sim \left \langle \frac{\bTb}{r^2 n} \bcdot \frac{\pd \bl}{\pd \ln r} \bigg|_{\rm twist} \right \rangle \sim \left \langle \frac{4\ag}{\cs^2} (r^2 n) \omb^2 \right \rangle.
\label{eq:dldt_visc_mag}
\ee
In the above estimate, we have assumed the relevant misalignment angles (between $\ld$ and $\blb$, or between $\ld$ and $\beb$) is of order unity.

\subsection{Formalism}
\label{sec:Form}

The torque per unit mass on the disk from the inner binary is given by Eq.~\eqref{eq:bTb}, with $\bTb = \bTb(r,t)$.  In addition, the gravitational potential from the binary induces a non-Keplarian angular frequency \citep{MirandaLai(2015)}, with
\be
\kappa^2 - n^2 = - 2 \omb n  f_{\rm b},
\label{eq:kappab}
\ee
where
\be
f_{\rm b} = \frac{1}{2} \left\{ \left[3 (\bl \bcdot \blb)^2 - 1\right] \left( 1 + \frac{3}{2} \eb^2 \right) - 15 \eb^2 (\bl \btimes \blb)^2\right\}.
\ee

When the Shakura-Sunaev $\ag$-viscosity parameter satisfies $\ag \lesssim H/r$, the disk lies in the bending wave regime \citep{PapaloizouLin(1995),LubowOgilvie(2000)}.  Any warp induced by an external torque is smoothed by bending waves passing through the disk.  Protoplanetary disks typically lie in the bending wave regime.  The time evolution of $\bl(r,t)$ is governed by the equations (\citealt{LubowOgilvie(2000)}; see also \citealt{Lubow(2002)})
\begin{align}
&\Sg r^2 n \frac{\pd \bl}{\pd t} = \frac{1}{r} \frac{\pd \bG}{\pd r} + \Sg \bTb,
\label{eq:dldt} \\
&\frac{\pd \bG}{\pd t} - \omb f_{\rm b} \bl \btimes \bG + \ag \Om \bG = \frac{\Sg \cs^2 r^3 n}{4} \frac{\pd \bl}{\pd r},
\label{eq:dGdt}
\end{align}
where $\bG$ is the internal torque.

From equation~\eqref{eq:tw}, we see that $t_{\rm bw} < \omb^{-1}$ for standard circumbinary disk parameters, so the disk should be only mildly warped.  We may therefore expand
\begin{align}
\bl(r,t) &= \ld(t) + \blone(r,t) + \dots, 
\label{eq:lexp} \\
\bG(r,t) &= \bG_0(r,t) + \bG_1(r,t) +  \dots
\end{align}
where $\ld$ is the unit vector along the total angular momentum of the disk, $|\blone| \ll |\ld| = 1$ [see Eqs.~\eqref{eq:lone_norm}-\eqref{eq:ld} below].  As we will see, the internal torque $\bG_0(r,t)$ maintains the rigid body dynamical evolution of $\ld$, while $\bG_1(r,t)$ maintains the warp profile $\blone$.  Perturbative expansions to study warped disk structure and time evolution have been taken by \cite{LubowOgilvie(2000),LubowOgilvie(2001)} and \cite{FoucartLai(2014)}.  Inserting~\eqref{eq:lexp} into Eq.~\eqref{eq:dldt}, integrating over $r \der r$, and using the zero torque boundary condition
\be
\bG_0(\rin,t) = \bG_0(\rout,t) = 0,
\label{eq:Gbdry}
\ee
we find the leading order time evolution of $\bl$ is given by
\be
\frac{\der \ld}{\der t} = -\bomb \left[ (1-\eb^2)(\ld \bcdot \blb) \blb \btimes \ld - 5 (\ld \bcdot \beb ) \beb \btimes \ld \right].
\label{eq:dl0dt}
\ee
Here,
\begin{align}
\bomb &= \frac{2\pi}{\Ld} \int_{\rin}^{\rout} \Sg r^3 \Om \omb \der r \simeq \frac{9 G \mub \ab^2}{16 \rin^2 \rout \sqrt{G \Mb \rout}}
\nonumber \\
&= 4.97 \times 10^{-5} \left( \frac{2 \ab}{\rin} \right)^2 \left( \frac{4 \mub}{\Mb} \right) 
\nonumber \\
&\hspace{15mm} \times \left( \frac{\Mb}{2 \, M_\odot} \right)^{1/2}  \left( \frac{\rout}{100 \, \text{AU}} \right)^{-3/2} \left( \frac{2\pi}{\text{yr}} \right)
\label{eq:bomb}
\end{align}
is the characteristic precession frequency of the rigid disk.  Equation~\eqref{eq:dl0dt} is equivalent to Equation~\eqref{eq:dldt_TP} if one replaces $\bomb$ with $\omb$, and the disk dynamics reduce to those of a test particle with $\bl = \ld$ when $\cs \to \infty$.

\subsection{Disk Warp Profile}
\label{sec:DiskWarp}

\begin{figure}
\centering
\includegraphics[scale=0.52]{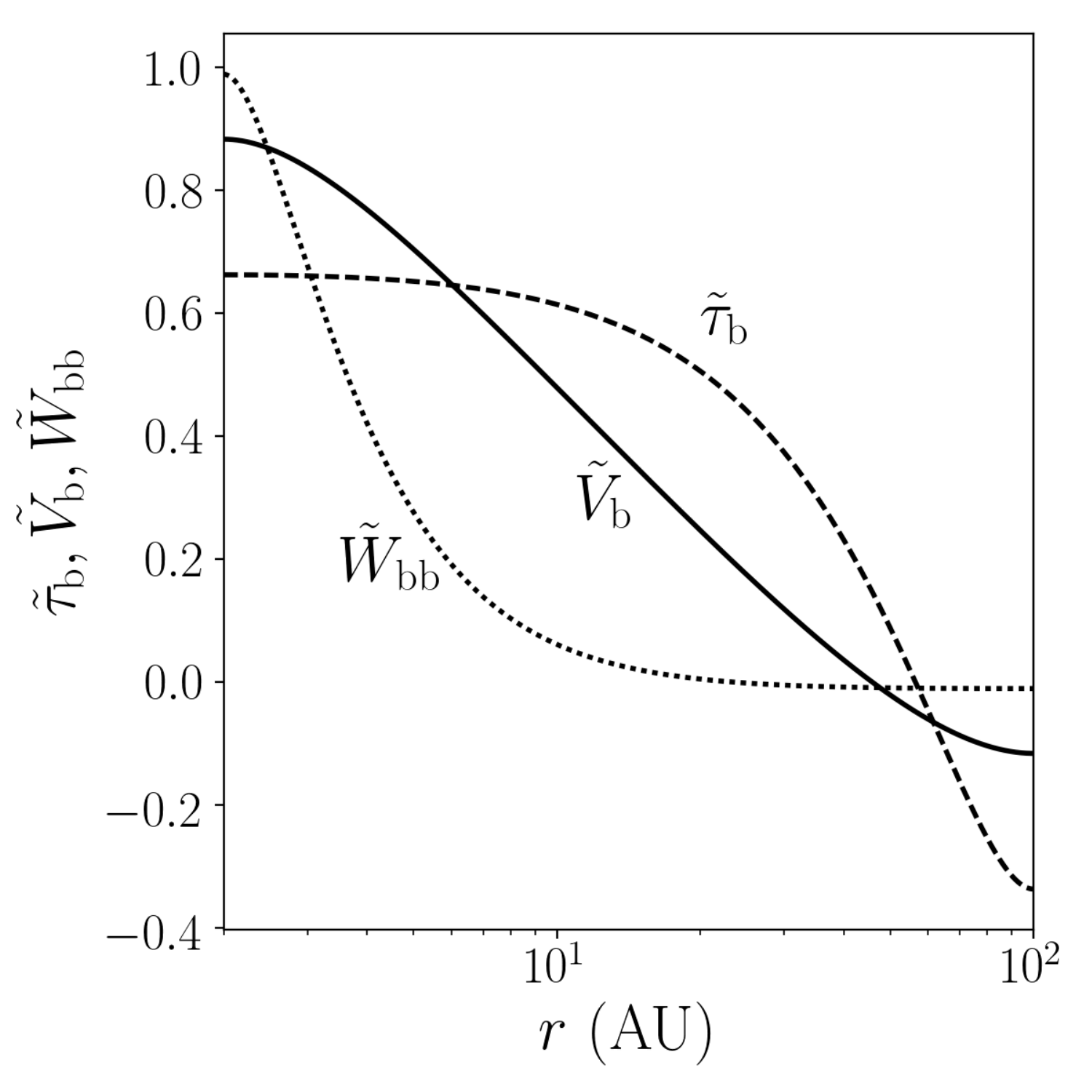}
\caption{Rescaled warp functions $\tilde \tau_{\rm b} = \tb/[\tb(\rin)-\tb(\rout)]$ [Eq.~\eqref{eq:tb}], $\tilde V_{\rm b} = \Vb/[\Vb(\rin)-\Vb(\rout)]$ [Eq.~\eqref{eq:Vb}], and $\tilde W_{\rm bb} = \Wbb/[\Wbb(\rin)-\Wbb(\rout)]$ [Eq.~\eqref{eq:Wbb}] as a function of radius.  We take $\rin = 2 \, \text{AU}$ and $\rout = 100 \, \text{AU}$.}
\label{fig:warp_funcs}
\end{figure}

With $\ld$ determined with boundary condition~\eqref{eq:Gbdry}, we may solve for $\bG_0(r,t)$:
\be
\bG_0(r,t) = \gb \left[ (1-\eb^2)(\ld \bcdot \blb) \blb \btimes \ld - 5 (\ld \bcdot \beb) \beb \btimes \ld \right],
\label{eq:G0}
\ee
where
\be
\gb(r) = \int_{\rin}^r \Sg {r'}^3 n (\omb - \bomb) \der r'.
\ee

With the leading order terms for $\bl$ and $\bG$, we may solve for $\blone$.  We impose the normalization condition
\be
\int_{\rin}^{\rout} \Sg r^3 \Om \blone(r,t) \der r = 0,
\label{eq:lone_norm}
\ee
so that $\ld$ is the unit vector along the \textit{total} angular momentum of the disk, or
\be
\ld(t) = \frac{2\pi}{\Ld} \int_{\rin}^{\rout} \Sg r^3 \Om \bl(r,t) \der r.
\label{eq:ld}
\ee
Inserting Eq.~\eqref{eq:G0} into Eq.~\eqref{eq:dGdt} and integrating, we obtain
\begin{align}
&\blone(r,t) = (\blone)_{\rm twist} + (\blone)_{\rm warp}
\nonumber \\
&+ 5 \bomb \gb(1-\eb^2) \ld \bcdot (\blb \btimes \beb) \left[ (\ld \bcdot \beb) \blb \btimes \ld - (\ld \bcdot \blb)\beb \btimes \ld \right],
\label{eq:lone}
\end{align}
where
\be
(\blone)_{\rm twist} = \Vb \left[ (1-\eb^2) (\ld \bcdot \blb)\blb \btimes \ld - 5(\ld \bcdot \beb)\beb \btimes \ld \right],
\label{eq:lone_visc}
\ee
and
\begin{align}
&(\blone)_{\rm warp} = 
\nonumber \\
&-\bomb \tb (1-\eb^2) (\ld \bcdot \blb)
\nonumber \\
\ &\times \left[ (1-\eb^2) (\ld \bcdot \blb) \blb \btimes (\blb \btimes \ld) - 5(\beb \bcdot \ld) \blb \btimes (\beb \btimes \ld) \right]
\nonumber \\
& + 5 \bomb \tb (\ld \bcdot \beb) 
\nonumber \\
\ &\times \left[ (1-\eb^2) (\ld \bcdot \blb) \beb \btimes (\blb \btimes \ld) - 5 (\beb \bcdot \ld)\beb \btimes (\beb \btimes \ld) \right]
\nonumber \\
&- \Wbb \fb 
\nonumber \\
\ &\times \left[ (1-\eb^2)(\ld \bcdot \blb)\ld \btimes (\blb \btimes \ld) - 5 (\beb \bcdot \ld) \ld \btimes (\beb \btimes \ld) \right].
\label{eq:lone_warp}
\end{align}
Here,
\begin{align}
\tb(r) &= \int_{\rin}^r \frac{\gb}{\Sg \cs^2 {r'}^3 n} \der r' - \tau_{\rm b0},
\label{eq:tb} \\
\Vb(r) &= \int_{\rin}^r \frac{\ag \gb}{\Sg \cs^2 {r'}^3} \der r' - V_{\rm b0},
\label{eq:Vb} \\
\Wbb(r) &= \int_{\rin}^r \frac{\omb \gb}{\Sg \cs^2 {r'}^3 n} \der r' - W_{\rm bb0},
\label{eq:Wbb}
\end{align}
and
\begin{align}
\tau_{\rm b0} &= \frac{2\pi}{\Ld}\int_{\rin}^{\rout} \Sg r^3 n \left( \int_{\rin}^r \frac{\gb}{\Sg \cs^2 {r'}^3 n} \der r' \right) \der r,
\label{eq:tb0} \\
V_{\rm b0} &= \frac{2\pi}{\Ld} \int_{\rin}^{\rout} \Sg r^3 n \left( \int_{\rin}^{r} \frac{\ag \gb}{\Sg \cs^2 {r'}^3} \der r' \right) \der r,
\label{eq:Vb0} \\
W_{\rm bb0} &= \frac{2\pi}{\Ld} \int_{\rin}^{\rout} \Sg r^3 n \left(\int_{\rin}^{r} \frac{\omb \gb}{\Sg \cs^2 {r'}^3 n} \der r' \right) \der r.
\label{eq:Wbb0}
\end{align}
The third term in Eq.~\eqref{eq:lone} arises from the fact that $\ld \bcdot \beb$ and $\ld \bcdot \blb$ are not constant in time, and is dynamically unimportant.  Although it is strait-forward  to compute the integrals in Eqs.~\eqref{eq:tb}-\eqref{eq:Wbb}, this calculation is tedious and unilluminating.  Instead, we notice that over most of the region in the integrals, the internal torque radial function $\gb(r)$ is of order
\be
\gb(r) \sim \Sg r^4 n \omb.
\label{eq:gb_approx}
\ee
Evaluating the warp functions and using the fact that $\rin \ll \rout$, we obtain the approximate expressions
\begin{align}
\bomb &\big[ \tb(\rin) - \tb(\rout) \big] \approx -0.108
\nonumber \\
&\times  \left( \frac{0.1}{h} \right)^2 \left( \frac{4 \mub}{\Mb} \right)^2 \left( \frac{2 \ab}{\rin} \right)^4 \left( \frac{50 \, \rin}{\rout} \right)^{3/2},
\label{eq:tb_scale} \\
\Vb&(\rin) - \Vb(\rout) \approx -0.258 
\nonumber \\
&\times \left( \frac{\ag}{0.01} \right) \left( \frac{0.1}{h} \right)^2 \left( \frac{4 \mub}{\Mb} \right) \left( \frac{2\ab}{\rin} \right)^2,
\label{eq:Vb_scale}\\
\Wbb&(\rin) - \Wbb(\rout) \approx -0.108
\nonumber \\
&\times  \left( \frac{0.1}{h} \right)^2  \left( \frac{4 \mub}{\Mb} \right)^2 \left( \frac{2 \ab}{\rin} \right)^4.
\label{eq:Wbb_scale}
\end{align}
  In Figure~\ref{fig:warp_funcs}, we plot the rescaled warp functions $\tilde \tau_{\rm b} = \tb/[\tb(\rin)-\tb(\rout)]$,  $\tilde V_{\rm b} = \Vb/[\Vb(\rin)-\Vb(\rout)]$, and $\tilde W_{\rm bb} = \Wbb/[\Wbb(\rin)-\Wbb(\rout)]$.

\subsection{Viscous Torques}
\label{sec:Visc}

\begin{figure*}
\includegraphics[scale=1.0]{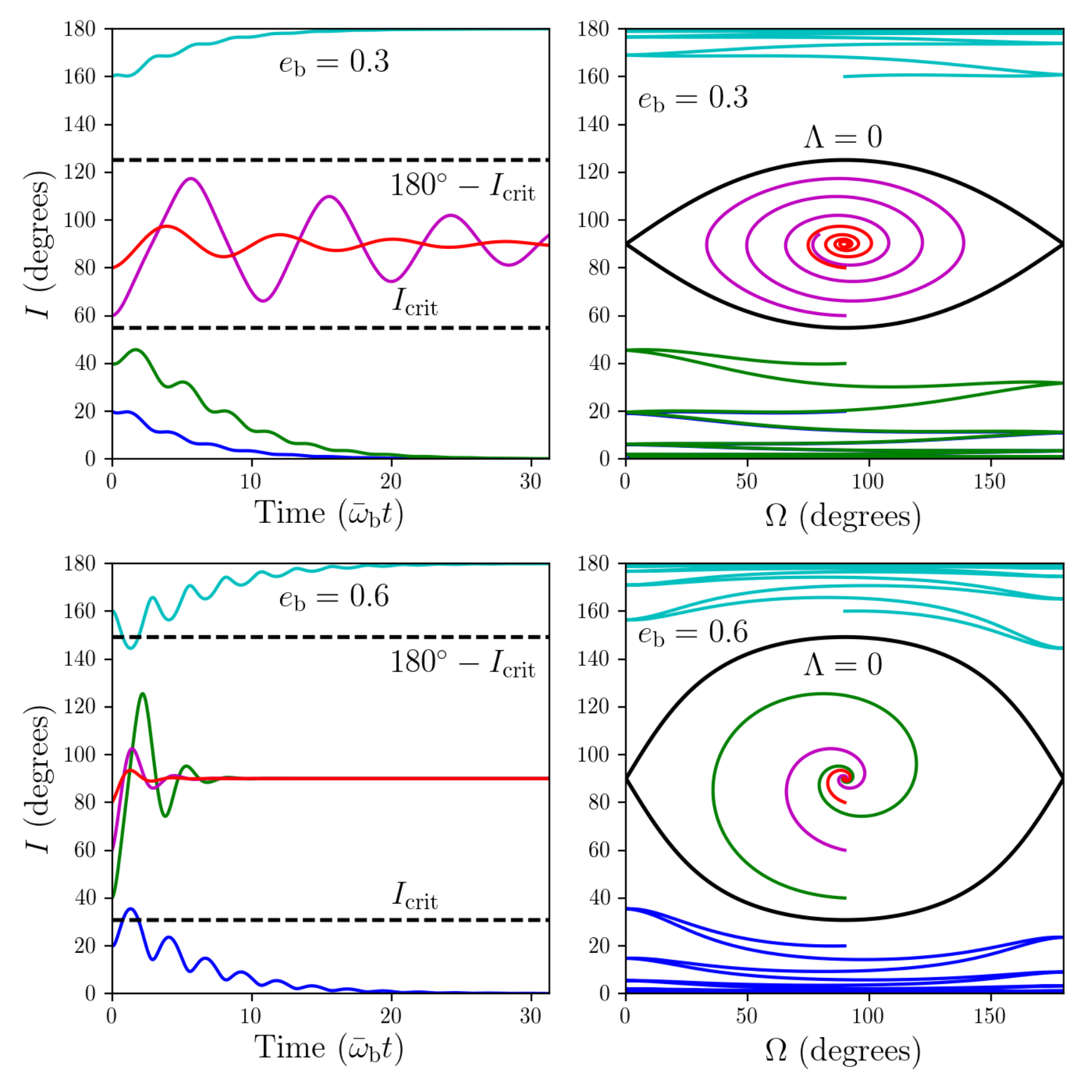}
\caption{ Time evolution of the disk orientation for two binary eccentricities $\eb$ as indicated.  {\bf Left Panels:} Disk inclination $I$ (the angle between $\ld$ and $\blb$) as a function of time.  The black dashed lines mark $\Icrit$ ($55^\circ$ for $\eb = 0.3$ and $31^\circ$ for $\eb = 0.6$) and $180^\circ-\Icrit$.  {\bf Right Panels:} Disk trajectories on the $I-\Om$ plane (where $\Om$ is the longitude of the ascending node of the disk).  The black solid curves mark the $\Lam = 0$ separatrix.  Initial values are $I(0) = 20^\circ$ (blue), $I(0) = 40^\circ$ (green), $I(0) = 60^\circ$ (magenta), $I(0) = 80^\circ$ (red), and $I(0) = 160^\circ$ (cyan), with $\Om(0) = 90^\circ$ for all trajectories.  The other parameters are $\Mb = 2 \, M_\odot$, $\mub = 0.5 \, M_\odot$, $\ab = 1 \, \text{AU}$, $\rin = 2 \, \text{AU}$, $\rout = 100 \, \text{AU}$, $\ag = 0.01$, and $h = 0.1$.}
\label{fig:visc}
\end{figure*}

The disk twisting due to viscosity $(\blone)_{\rm twist}$ [Eq.~\eqref{eq:lone_visc}] interacts with $\bTb$ [Eq.~\eqref{eq:bTb}], effecting the evolution of $\ld$ over viscous timescales.  Inserting $\bl = \ld + (\blone)_{\rm twist}$ into Equation~\eqref{eq:dldt}, integrating over $2\pi r \der r$, and using the boundary condition~\eqref{eq:Gbdry}, we obtain
\begin{align}
 \left( \frac{\der \bLd}{\der t} \right)_{\rm visc} = & \ \Ld \cgb \Big[(1-\eb^2) (\ld \bcdot \blb)^2 \blb \btimes (\blb \btimes \ld)
\nonumber \\
&+ 25 (\ld \bcdot \beb)^2 \beb \btimes (\beb \btimes \ld)
\nonumber \\
&- 5 (1-\eb^2) (\ld \bcdot \beb)(\ld \bcdot \blb) \blb \btimes (\beb \btimes \ld)
\nonumber \\
&- 5 (1-\eb^2)(\ld \bcdot \beb)(\ld \bcdot \blb)\beb \btimes (\blb \btimes \ld) \Big].
\label{eq:dLdt_visc}
\end{align}
Here,
\be
\cgb = \frac{2\pi}{\Ld} \int_{\rin}^{\rout} \frac{4 \ag \gb^2}{\Sg \cs^2 r^3} \der r .
\label{eq:cgb}
\ee
Using the approximate expression of $\gb$ [Eq.~\eqref{eq:gb_approx}], one may easily reproduce Eq.~\eqref{eq:dldt_visc_mag}. The same argument used in the calculation of Eqs.~\eqref{eq:tb_scale}-\eqref{eq:Wbb_scale} may be used to calculate the approximate expression of the viscous rate $\cgb$:
\begin{align}
\cgb \approx \ & 1.02 \times 10^{-5} \left( \frac{\ag}{0.01} \right) \left( \frac{0.1}{h} \right)^2 \left( \frac{2\ab}{\rin} \right)^4
\nonumber \\
&\times  \left( \frac{4\mub}{\Mb} \right)^2 \left(\frac{\Mb}{2 \, M_\odot} \right)^{1/2} \left( \frac{100 \, \text{AU}}{\rout} \right)^{3/2} \left( \frac{2\pi}{\text{yr}} \right).
\label{eq:cgb_scale}
\end{align}
{ We choose to normalize $\cgb$ by $\ag = 0.01$; real protoplanetary disks may have $\ag$ in the range $10^{-1} - 10^{-5}$ \citep{Rafikov(2017)}.  } Since
\be
\frac{\der \ld}{\der t} = \frac{1}{\Ld} \left( \frac{\der \bLd}{\der t} - \frac{\der \Ld}{\der t} \ld \right),
\ee
the viscous dissipation from disk twisting effects the evolution of $\ld$ according to
\begin{align}
\bigg( \frac{\der \ld}{\der t} \bigg)_{\rm visc}  = \cgb \Lam \big[  &(1-\eb^2) (\ld \bcdot \blb) \ld \btimes (\blb \btimes \ld) 
\nonumber \\
&- 5  (\ld \bcdot \beb) \ld \btimes (\beb \btimes \ld) \big],
\label{eq:dldt_visc}
\end{align}
where $\Lam$ is given by Eq.~\eqref{eq:Lam}, except we replace $\bl$ by $\ld$:
\be
\Lam = (1-\eb^2)(\ld \bcdot \blb)^2 - 5 (\ld \bcdot \beb)^2.
\label{eq:Lamd}
\ee

Equation~\eqref{eq:dldt_visc} is the main result of our technical calculation.  We see
\begin{align}
\frac{\der}{\der t}(\blb \bcdot \ld) \bigg|_{\rm visc} &= \cgb \Lam (\ld \bcdot \blb) \big[ (1-\eb^2) - \Lam \big],
\label{eq:dcosdbdt} \\
\frac{\der}{\der t}(\beb \bcdot \ld) \bigg|_{\rm visc} &= - \cgb \Lam (\ld \bcdot \beb) \big[ \Lam + 5 \eb^2 \big],
\label{eq:dcosdedt}
\end{align}
Because $-5\eb^2 < \Lam < (1-\eb^2)$ [Eq.~\eqref{eq:Lamd}], Equations~\eqref{eq:dcosdbdt}-\eqref{eq:dcosdedt} show the system has two different end-states depending on the initial value for $\Lam$:
\begin{enumerate}
\item $\Lam > 0$: The viscous torque \eqref{eq:dldt_visc} pushes $\ld$ towards $\blb$.  The final state of $\ld$ is alignment (if $\blb \bcdot \ld > 0$ initially) or anti-alignment (if $\blb \bcdot \ld < 0$ initially) with $\blb$.
\item $\Lam < 0$: The viscous torque \eqref{eq:dldt_visc} pushes $\ld$ towards $\beb$.  The final state of $\ld$ is alignment (or anti-alignment) with $\beb$.
\end{enumerate}

Figure~\ref{fig:visc} shows several examples of the results for the evolution of disk orientation, obtained by integrating the time evolution of $\ld$, including gravitational [Eq.~\eqref{eq:dl0dt}] and viscous [Eq.~\eqref{eq:dldt_visc}] torques.  On the left panels, we plot the disk inclination $I$ with time, for the binary eccentricities indicated.  We choose the initial $\Om(0) = 90^\circ$ for all cases, so that $I < \Icrit$ ($I > \Icrit$) corresponds exactly to $\Lam > 0$ ($\Lam < 0$) (see Eqs.~\eqref{eq:Lam} and~\eqref{eq:Icrit}).  Thus we expect $I \to 0^\circ$  when $I < \Icrit$, $I \to 90^\circ$ when $\Icrit < I < 180^\circ - \Icrit$, and $I \to 180^\circ$ when $I > 180^\circ - \Icrit$.  On the right panels of Figure~\ref{fig:visc}, we plot the disk trajectories on the $I-\Om$ plane [Eq.~\eqref{eq:l_coos} with $\bl \to \ld$].  Again, we see when $I < \Icrit$ ($\Lam > 0$), $\ld$ aligns with $\blb$, while when $I > \Icrit$ ($\Lam < 0$), $\ld$ aligns with $\beb$, as expected.

\section{Secular Dynamics with Massive Inclined Outer Body}
\label{sec:ArbJ}

\begin{figure*}
\includegraphics[scale=0.53]{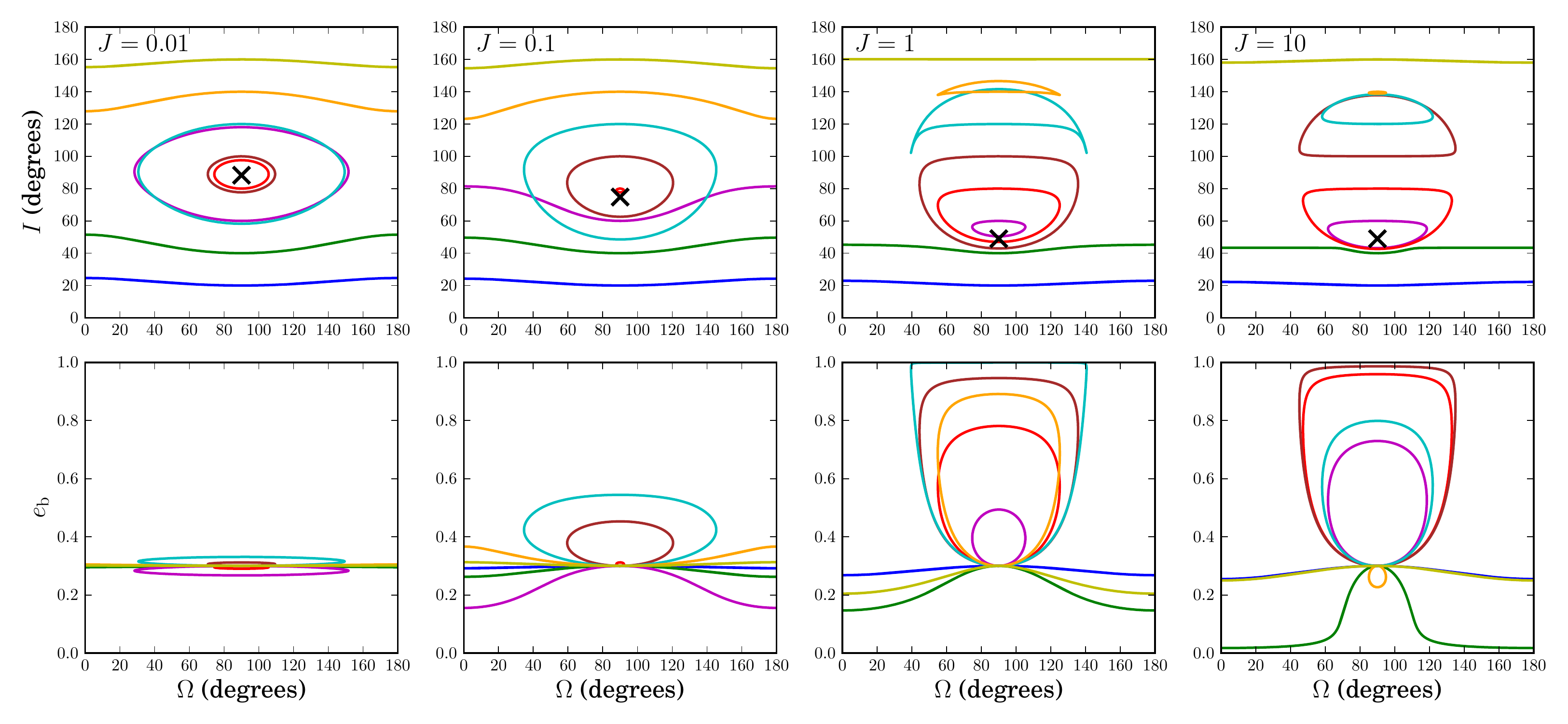}
\caption{{ Trajectories in the $I-\Om$ (top panels) and $\eb-\Om$ (bottom panels) planes for the $J$ values indicated, with fixed points $(I,\Om) = (I_{\rm fp},90^\circ)$ computed with Eq.~\eqref{eq:FP} marked with black x's.  Initial values for the trajectories are $I(0) = 20^\circ$ (blue), $I(0) = 40^\circ$ (green), $I(0) = 60^\circ$ (magenta), $I(0) = 80^\circ$ (red), $I(0) = 100^\circ$ (brown), $I(0) = 120^\circ$ (cyan), $I(0) = 140^\circ$ (orange), and $I(0) = 160^\circ$ (yellow), with $\Om(0) = 90^\circ$ and $\eb(0) = 0.3$ for all trajectories.} }
\label{fig:DynArbJ}
\end{figure*}

\begin{figure}
\includegraphics[scale=0.55]{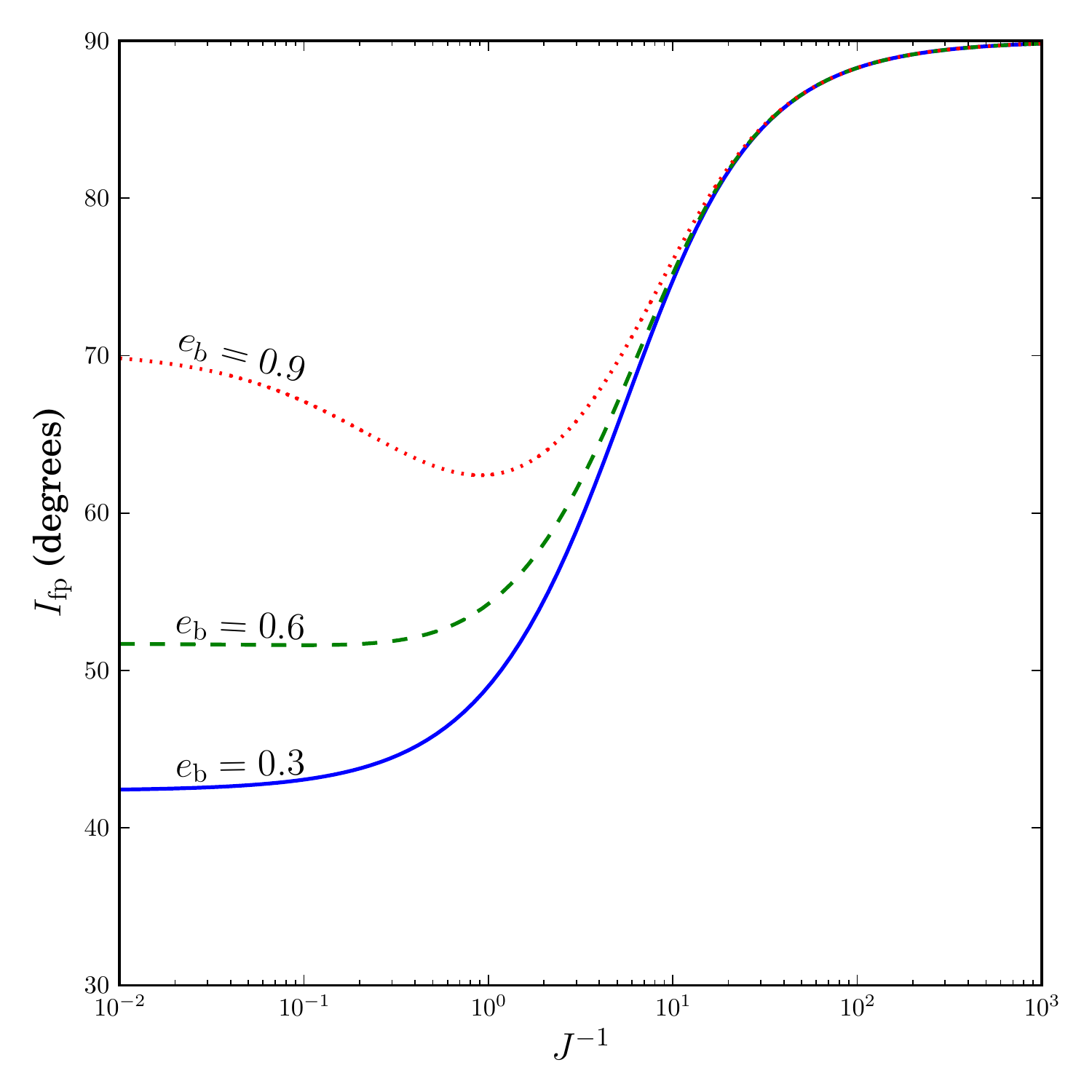}
\caption{{ Inclination $I_{\rm fp}$ as a function of $J^{-1}$, computed with Eq.~\eqref{eq:FP} with $\Om = \pi/2$.  The binary eccentricity $\eb = \eb(0)$ takes values as indicated.} }
\label{fig:FixPointArbJ}
\end{figure}

{
Sections~\ref{sec:TestPartDyn}-\ref{sec:CircBinDiskDyn} neglected the circumbinary disk's angular momentum, a valid assumption as long as $\Ld \ll L_{\rm b}$ [Eq.~\eqref{eq:LdOverLb}].  When $\Ld \gtrsim L_{\rm b}$, the non-zero disk angular momentum will change the locations of the fixed points of the system, and hence may effect its dynamical evolution over viscous timescales.
}

{
Consider the setup of Section~\ref{sec:TestPartDyn}, except we now include the outer body's mass $m$ and angular momentum ${\bm L} = m \sqrt{G \Mb r} \bl$.  The evolution equations for $\bl$, $\bjb = \sqrt{1-\eb^2} \blb$, and $\beb$ are [\citealt{Liu(2015)}; Eqs. (17)-(19)]
}
\begin{align}
\frac{\der \bl}{\der t} &= - \omb \left[ (\bjb \bcdot \bl) \bjb \btimes \bl - 5 (\beb \bcdot \bl) \beb \btimes \bl \right], \\
\frac{\der \bjb}{\der t} &= J \omb \left[ (\bjb \bcdot \bl) \bjb \btimes \bl - 5 (\beb \bcdot \bl) \beb \btimes \bl \right], \\
\frac{\der \beb}{\der t} &= J \omb \left[ (\bjb \bcdot \bl) \beb \btimes \bl + 2 \bjb \btimes \beb - 5 (\beb \bcdot \bl) \bjb \btimes \bl \right],
\end{align}
{
where
}
\begin{align}
J &= \frac{L}{L_{\rm b}/\sqrt{1-\eb^2}} = \frac{\mu}{\mub} \left( \frac{\Mb+m}{\Mb} \right)^{1/2} \left( \frac{r}{\ab} \right)^{1/2},
\label{eq:J} \\
\omb &= \frac{3}{4} \left( \frac{m}{\mu} \right) \left( \frac{\mub}{\Mb} \right) \left( \frac{\Mb}{\Mb + m} \right)^{1/2} \left( \frac{\ab}{r} \right)^{7/2} \sqrt{ \frac{G \Mb}{\ab^3} },
\label{eq:omb_J}
\end{align}
{
and $\mu = m \Mb/(m+\Mb)$.  Equation~\eqref{eq:omb_J} reduces to Eq.~\eqref{eq:omb} when $m \to 0$.  The conservations of total quadrupole potential energy [see Eq.~\eqref{eq:Phi_quad}] and total angular momentum yield two constants of motion (e.g. \citealt{Liu(2015),Anderson(2017)})
}
\begin{align}
\Psi &= 1 - 6 \eb^2 - 3(1-\eb^2) \cos^2 I + 15 \eb^2 \sin^2 I \sin^2 \Om,
\label{eq:Psi} \\
K &= \sqrt{1-\eb^2} \cos I - \frac{\eb^2}{2 J}.
\label{eq:K}
\end{align}
{
For a given $K$, one may solve Eq.~\eqref{eq:K} to get $\eb^2 = \eb^2(I)$.  Assuming $0 \le I \le \pi/2$ and requiring $0 \le \eb < 1$, we obtain
}
\be
\eb^2 = 2 J^2 \left[ \cos I \sqrt{ \left( \frac{2K}{J} + \frac{1}{J^2} \right) + \cos^2 I } - \left( \frac{K}{J} + \cos^2 I \right) \right].
\label{eq:eb2}
\ee
{
Equation~\eqref{eq:Psi} then gives $\Psi = \Psi(I,\Om)$.  When $J \sim K^{-1} \ll 1$, Eq.~\eqref{eq:eb2} reduces to
}
\be
\eb^2 \simeq -2 K J = \text{constant},
\ee
{
while when $J \gg 1$, Eq.~\eqref{eq:eb2} becomes
}
\be
\eb^2 \simeq 1 - \frac{K^2}{\cos^2 I}.
\ee

{
The fixed points of the system in the $I-\Om$ plane are determined by
}
\be
\frac{\pd \Psi}{\pd I} = \frac{\pd \Psi}{\pd \Om} = 0.
\label{eq:FP}
\ee
{
The condition $\pd \Psi/\pd \Om = 0$ gives $\Om = \pi/2$ and $\Om = 3\pi/2$, as before (see Sec.~\ref{sec:TestPartDyn}).  For arbitrary $J$, one must numerically solve
$\pd \Psi/\pd I|_{\Om = \pi/2, 3\pi/2} = 0$
to calculate the fixed points $I = I_{\rm fp} > 0$ ($I = 0$ is always a fixed point of the system).  However, when $J \ll 1$, one may show analytically that (as found in Sec.~\ref{sec:TestPartDyn})
}
\be
I_{\rm fp} \simeq \pi/2,
\label{eq:Ifp_pol}
\ee
{
while when $J \gg 1$,
}
\be
I_{\rm fp} \simeq \cos^{-1} \sqrt{ \frac{3(1-\eb^2)}{5} },
\label{eq:Ifp_LK}
\ee
{
where $\eb^2 = \eb^2(0)$. Notice $I_{\rm fp}$ is the Lidov-Kozai critical inclination when $J \gg 1$ and $\eb(0) = 0$ \citep{Lidov(1962),Kozai(1962)}.
}

{
Figure~\ref{fig:DynArbJ} plots trajectories of the system in the $I-\Om$ and $\eb-\Om$ planes.  When $J \ll 1$, the system's dynamics reduce to that discussed in Section~\ref{sec:TestPartDyn}, with $I_{\rm fp} \simeq 90^\circ$ (black x's), $\eb \simeq \eb(0)$,  and trajectories above and below $I = 90^\circ$ are symmetric.  As $J$ increases in magnitude, $I_{\rm fp}$ decreases, $\eb$ begins to oscillate, and the inclination symmetry above and below $I = 90^\circ$ is lost.  Although different trajectories may cross in the $I-\Om$ plane, each still has a unique $\Psi$ value [Eq.~\eqref{eq:Psi}], since the binary's $\eb$ value differs from Eq.~\eqref{eq:eb2} when $I > \pi/2$.  When $J \gg 1$, the system's dynamics approaches the classic Lidov-Kozai regime \citep{Lidov(1962),Kozai(1962)}.  The fixed point $I_{\rm fp}$ of the system approaches Eq.~\eqref{eq:Ifp_LK}, with $\eb$ reaching large values when $I(0) > I_{\rm fp}$, and with trajectories symmetric above and below $I = 90^\circ$.
}

{
Figure~\ref{fig:FixPointArbJ} plots $I_{\rm fp}$ as a function of $J^{-1}$, computed for $\Om = \pi/2$ with the $\eb = \eb(0)$ values as indicated.  The two limiting cases given by Eqs.~\eqref{eq:Ifp_pol} and~\eqref{eq:Ifp_LK} are achieved when $J \ll 1$ and $J \gg 1$, respectively, and $I_{\rm fp}$ generally varies non-monitonically with increasing $J$.  Since $\eb$ should evolve in time under the influence of viscous dissipation from disk warping, one cannot determine the final value of $I_{\rm fp}$ the system may evolve into 
starting from initial $I(0)$ and $\eb(0)$ values without a detailed calculation similar to Section~\ref{sec:CircBinDiskDyn}.  Nevertheless, Figures~\ref{fig:DynArbJ} and~\ref{fig:FixPointArbJ} show there exist highly inclined fixed points for any value of $J$.
For $J \lesssim 0.1$, the system may evolve into near polar alignment, with $I_{\rm fp}$ somewhat less than $90^\circ$.
}

%%%%%%%%%%%%%%%%%%%%%%%%%%%%%%%%%%%
\section{Torque on Binary and Effect of Accretion}
\label{sec:TorqueBinAcc}

In the previous sections, we have studied the evolution of the disk around a binary with fixed $\blb$ and $\beb$.  Here we study the back-reaction torque on the binary from the disk.  First consider a circular binary.  The viscous back reaction torque on the binary from the disk is [Eq.~\eqref{eq:dLdt_visc}]
\begin{align}
\left( \frac{\der {\bm L}_{\rm b}}{\der t} \right)_{\rm visc} &= - \left( \frac{ \der \bLd}{\der t} \right)_{\rm visc}  \nonumber \\
&= -\Ld \cgb(\ld \bcdot \blb)^2 \blb \btimes (\blb \btimes \ld).
\label{eq:dLbdt_visc}
\end{align}
In addition, accretion onto the binary from the disk adds angular momentum to the binary's orbit:
\be
\left( \frac{\der \bLb}{\der t} \right)_{\rm acc} \simeq \lam \dot M \sqrt{G \Mb \rin} \ld.
\label{eq:dLbdt_acc}
\ee
Here, $\dot M$ is the mass accretion rate onto the binary, $\lam \sim 1$ (e.g. \citealt{Miranda(2017)}), and we have assumed $\bl(\rin,t) \simeq \ld(t)$ (see below).  The torques~\eqref{eq:dLbdt_visc} and~\eqref{eq:dLbdt_acc} are equivalent to those considered in \cite{FoucartLai(2013)}, except we give different power-law prescriptions for $\Sg = \Sg(r)$ and $H = H(r)$.  For disks in steady state, we have
\be
\dot M \simeq 3 \pi \ag h^2 \Sgin \rin^2 n(\rin),
\ee
Using Eqs.~\eqref{eq:dLdt_visc} (with $\eb = 0$), \eqref{eq:dLbdt_visc} and~\eqref{eq:dLbdt_acc}, we obtain the net disk-binary alignment timescale for small angle between $\ld$ and $\blb$:
\be
t_{\rm align} = \cgb^{-1} \left[ 1 + (1+\eta) \frac{\Ld}{L_{\rm b}} \right]^{-1},
\label{eq:t_align}
\ee
where
\begin{align}
\eta &\equiv \frac{\lam \dot M \sqrt{G \Mb \rin}}{\Ld \cgb}
\nonumber \\
&\approx 0.031 \, \lam \left( \frac{h}{0.1} \right)^4 \left( \frac{\rin}{2 \ab} \right)^4 \left( \frac{\Mb}{4\mub} \right)^2
\end{align}
measures the strength of the accretion torque to the viscous torque on the binary ($\eta/\lam = f^{-1}$, $\lam = g$ in the notation of \citealt{FoucartLai(2013)}).

Since $\bl(\rin,t) \ne \bl(\rout,t)$, the disk angular momentum loss through accretion causes $\bl_{\rm d}$ to change with time:
\be
\bigg( \frac{\der \bl_{\rm d}}{\der t} \bigg)_{\rm acc} \sim -\frac{\lam \dot M \sqrt{ G \Mb \rin}}{\Ld} \Big\{ \bl(\rin,t) - \bl_{\rm d} \big[ \bl_{\rm d} \bcdot \bl(\rin,t) \big] \Big\}.
\ee
Because the magnitude of the tilt of $\bl(\rin,t)$ from $\bl_{\rm d}$ is of order 
\be
\big[ \bl(\rin,t) - \bl_{\rm d} \big] \sim - \frac{\pd \bl}{\pd \ln r} \bigg|_{\rm warp},
\ee
we find
\be
\bigg( \frac{\der \bl_{\rm d}}{\der t} \bigg)_{\rm acc} \sim -\frac{4 \lam \dot M \sqrt{G \Mb \rin}}{\cs^2 \Ld} \left( \frac{\kg^2 - n^2}{2 n^2} \right) \bl_{\rm d} \btimes \bTb.
\label{eq:dldt_acc_mag}
\ee
Detailed calculation shows that the accretion torque~\eqref{eq:dldt_acc_mag} is always much less than the viscous torque~\eqref{eq:dldt_visc_mag} on the disk.  We relegate the calculation and discussion of the accretion torque~\eqref{eq:dldt_acc_mag} to the Appendix.

For eccentric binaries, the back-reaction toque from the disk is
$\der \bLb/ \der t = - \der \bLd/ \der t$ [Eq.~\eqref{eq:dLdt_visc}]. But this is not sufficient for determining the evolution of $\beb$ and $\blb$.  In addition, how accretion affects the binary eccentricity is also uncertain
(e.g. \citealt{Rafikov(2016),Miranda(2017)}). Nevertheless, as long as $L_{\rm b} \gtrsim L_{\rm d}$, the timescale for the disk-binary inclination evolution should be of order $\cgb^{-1}$, with an estimate given by Eq.~\eqref{eq:cgb_scale}.

\section{Discussion}
\label{sec:Discuss}

\subsection{Theoretical Uncertainties}
\label{sec:TheoryUncertain}

Our theoretical analysis of disks around binaries assumes a linear disk warp.  However, we find that at the inner disk region, $|\pd \bl/\pd \ln r|$ reaches $\sim 0.1$ for a wide range of binary and disk parameters.  Inclusion of weakly non-linear warps in Equations~\eqref{eq:dldt}-\eqref{eq:dGdt} may introduce new features in the disk warp profile \citep{Ogilvie(2006)}.  In addition, disk warps of this magnitude may interact resonantly with inertial waves in the disk, leading to a parametric instability which may excite turbulence in the disk \citep{Gammie(2000),OgilvieLatter(2013)}.  An investigation of these effects is outside the scope of this paper, but their inclusion is unlikely to change the direction of disk-binary inclination evolution (alignment vs polar alignment).

\subsection{Observational Implications}
\label{sec:ObsImps}

\begin{figure}
\includegraphics[scale=0.8]{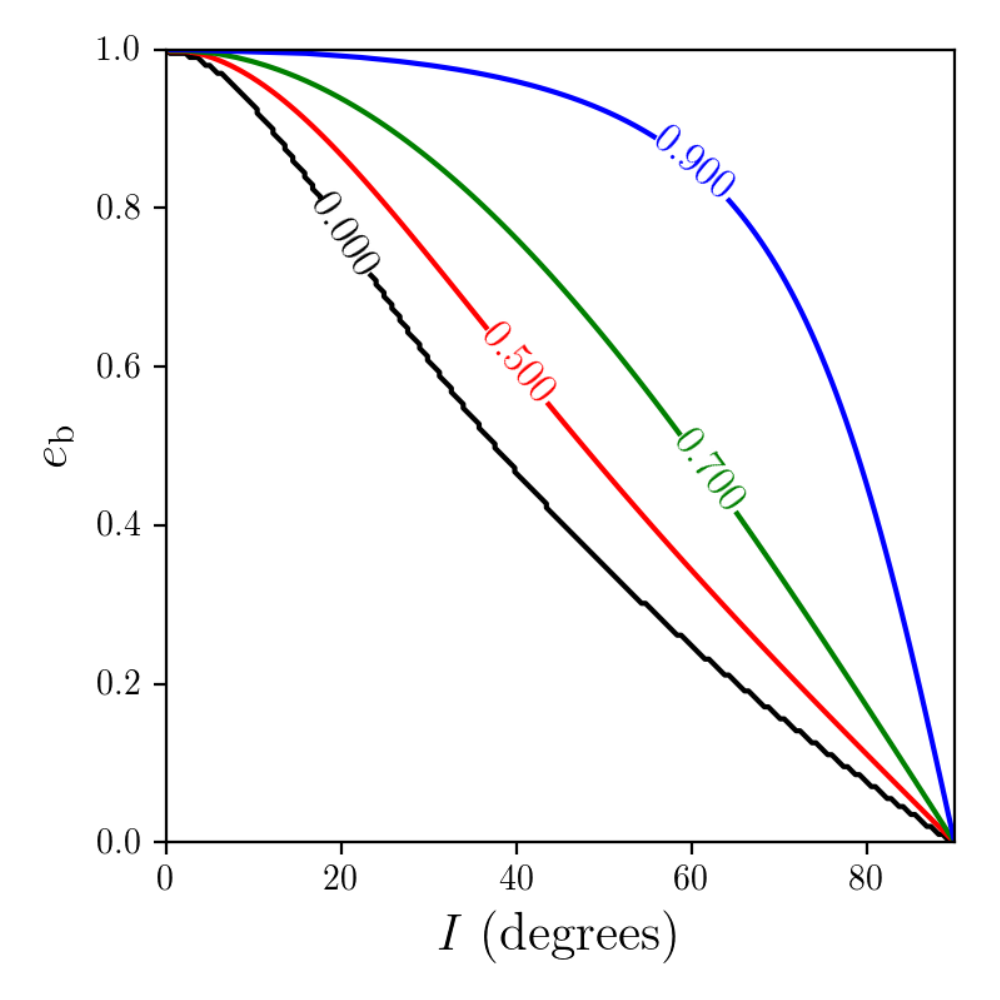}
\caption{Contour plot of the probability of polar alignment $P_{\rm polar}$ [Eq.~\eqref{eq:P_polar}] as a function of disk inclination $I$ and binary eccentricity $\eb$.  Contours of constant $P_{\rm polar}$ are labeled as indicated.  The $P_{\rm polar} = 0$ line (black) traces out $\Icrit$ [Eq.~\eqref{eq:Icrit}], while the $P_{\rm polar} = 0.5$ line (red) traces out $I_{\rm polar}$ [Eq.~\eqref{eq:Ipolar}].}
\label{fig:P_polar}
\end{figure}

In Section~\ref{sec:Visc}, we showed that the viscous torque associated with disk twist/warp tends to drive the circumbinary disk axis $\ld$ toward $\pm \blb$ (alignment or anti-alignment) when $\Lam > 0$, and toward $\pm \beb$ (polar alignment) when $\Lam < 0$.  Note that $\Icrit < I < 180^\circ - \Icrit$ is a necissary, but not sufficient condition for polar alignment of the disk [Eq.~\eqref{eq:Icrit}].  An extreme example is when $\Om = 0^\circ$, since $\Lam \ge 0$ for all inclinations $I$.  Because the circumbinary disk probably formed in a turbulent molecular cloud, the disk is unlikely to have a preferred $\Om$ when it forms.  The condition for polar alignment ($\Lam < 0$) requires $\Om$ to satisfy
%DL: 
%\be
%\sin^2 \Om > \frac{1}{5} \left( \frac{1}{\eb^2} - 1 \right) \cot^2 I.
%\ee
\be
\sin^2 \Om > \left(\frac{1-\eb^2}{5\eb^2}\right) \frac{1}{\tan^2 I} = \frac{\tan^2 I_{\rm crit}}{ \tan^2 I}.
\ee
%DL:
Assuming a uniform distribution of $\Om$-values from $0$ to $2\pi$, the probability of the disk to
polar align is (for given $I,\eb$)
\be
P_{\rm polar}(I,\eb) = 1 - 2\Om_{\rm min}/\pi.
\label{eq:P_polar}
\ee
where
\be
\Om_{\rm min}(I,\eb) =
\arraycolsep=1.4pt\def\arraystretch{2.2}
\left\{ \begin{array}{cc}
\pi/2 &  |\sin I| \le | \sin \Icrit |  \\
%DL:
%\sin^{-1} \sqrt{ \frac{1}{5} \left( \eb^{-2} - 1 \right) \cot^2 I } \hspace{1mm} & I > \Icrit
\sin^{-1} \left( \frac{ \tan\Icrit }{ |\tan I| } \right) \hspace{1mm} & {\rm otherwise}
\end{array}
\right. 
\ee
%we see the disk evolves to be polar to the binary orbital plane when
%$\sin^2 \Om > \sin^2 \Om_{\rm min}$.  Assuming a uniform distribution
%of $\Om$-values over $0$ to $2\pi$, the probability of the disk to polar align is
%When $P_{\rm polar} > 0.5$, alignment of the circumbinary disk with
%the binary orbital plane becomes unlikely.  
We define the inclination $I_{\rm polar}$ through $P_{\rm polar}(I_{\rm polar},\eb)~=~0.5$.
Solving for $I_{\rm polar}$, we obtain
\be
%I_{\rm polar} = \tan^{-1} \sqrt{ 2(\eb^{-2}-1)/5 }
I_{\rm polar} = \tan^{-1} \sqrt{ 2(1-\eb^2)/5\eb^2 }
\label{eq:Ipolar}
\ee
%DL:
In Figure~\ref{fig:P_polar}, we plot contours of constant $P_{\rm polar}$ 
in the $I-\eb$ space.  The $P_{\rm polar} = 0$
curve (black) traces out $I_{\rm crit}$ [Eq.~\eqref{eq:Icrit}], while
the $P_{\rm polar} = 0.5$ curve (red) traces out $I_{\rm polar}$
[Eq.~\eqref{eq:Ipolar}].  When $I < \Icrit$, alignment of $\bl$ with
$\blb$ is inevitable.  When $I > I_{\rm polar}$, alignment of $\bl$
with $\beb$ is probable.

\begin{table}
\centering
\begin{tabular}{ |l|c|c|c| }
\hline
Binary & $\eb$ & $I_{\rm crit}$ & $I_{\rm polar}$ \\
\hline
99 Herculis & 0.77 & $20^\circ$ & $28^\circ$ \\
$\ag$ CrB & 0.37 & $48^\circ$ & $58^\circ$ \\
$\beta$ Tri & 0.43 & $43^\circ$ & $53^\circ$ \\
DQ Tau & 0.57 & $33^\circ$ & $42^\circ$ \\
AK Sco & 0.47 & $40^\circ$ & $50^\circ$ \\
HD 98800 B & 0.78 & $20^\circ$ & $27^\circ$ \\
\hline
\end{tabular}
\caption{Binary eccentricities $\eb$, with their inclinations $\Icrit$ [Eq.~\eqref{eq:Icrit}] and $I_{\rm polar}$ [Eq.~\eqref{eq:Ipolar}], for the selected eccentric binaries with circumbinary disks.  With the exception of the debris disk around 99 Herculis, all binaries have circumbinary disks aligned with the binary orbital plane within a few degrees.  Binary eccentricities are from \protect\cite{Kennedy(2012a)} (99 Herculis), \protect\cite{TomkinPopper(1986)} ($\alpha$ CrB), \protect\cite{Pourbaix(2000)} ($\bg$ Tri), \protect\cite{Czekala(2016)} (DQ Tau), \protect\cite{Alencar(2003)} (AK Sco), and \protect\cite{Boden(2005)} (HD 98800 B)
}
\label{tab:binary}
\end{table}

Table~\ref{tab:binary} lists a number of circumbinary systems with highly eccentric binaries.  With the exception of 99 Herculis, all the binaries listed have disks coplanar with the binary orbital plane within a few degrees.  We also list $\Icrit$ [Eq.~\eqref{eq:Icrit}] and $I_{\rm polar}$ [Eq.~\eqref{eq:Ipolar}] for these systems.  { We do not list the binaries KH 15D \citep{Winn(2004),ChiangMurray-Clay(2004),Capelo(2012)} and HD 142527B \citep{Marino(2015),Casassus(2015),Lacour(2016)} since the orbital elements of these binaries are not well constrained.  However, both binaries appear to have significant eccentricities \citep{ChiangMurray-Clay(2004),Lacour(2016)}.}

Since planets form in gaseous circumbinary disks, planets may form with orbital planes perpendicular to the binary orbital plane if the binary is sufficiently eccentric.  Such planets may be detectable in transit surveys of eclipsing binaries due to nodal precession of the planet's orbit.

The twist and warp calculated in Section~\ref{sec:DiskWarp} is non-negligible.  Further observations of (gaseous) circumbinary disks may be able to detect such warps \citep{JuhaszFacchini(2017)}, further constraining the orientation and dynamics of circumbinary disk systems.

\section{Summary}
\label{sec:Summary}

Using semi-analytic theory, we have studied the warp and long-term evolution of circumbinary disks around eccentric binaries.  Our main results and conclusions are listed below.
\begin{enumerate}
\item For protoplanetary disks with dimensionless thickness $H/r$ larger than the viscosity parameter $\ag$, bending wave propagation effectively couples different regions of the disk, making it precess as a quasi-rigid body.  Without viscous dissipation from disk warping, the dynamics of such a disk is similar to that of a test particle around an eccentric binary (Secs.~\ref{sec:TestPartDyn} and~\ref{sec:Form}).
\item When the binary is eccentric and the disk is significantly inclined, the disk warp profile exhibits new features not seen in previous works.  The disk twist [Eq.~\eqref{eq:lone_visc}] and warp [Eq.~\eqref{eq:lone_warp}] have additional contributions due to additional torques on the disk when the binary is eccentric.
\item Including the dissipative torque from warping, the disk may evolve to one of two states, depending on the initial sign of $\Lam$ [Eq.~\eqref{eq:Lam}] (Sec.~\ref{sec:Visc}).  When $\Lam$ is initially positive, the disk angular momentum vector aligns (or anti-aligns) with the binary orbital angular momentum vector.  When $\Lam$ is initially negative, the disk angular momentum vector aligns with the binary eccentricity vector (polar alignment).  Note that $\Lambda$ depends on both $I$ (the disk-binary inclination) and $\Omega$ (the longitude of ascending node of the disk).  Thus for a given $\eb$, the direction of inclination evolution depends not only on the initial $I(0)$, but also on the initial $\Om(0)$.
\item { When the disk has a non-negligible angular momentum compared to the binary, 
the system’s fixed points are modified (Sec.~\ref{sec:ArbJ}). The disk may then evolve to a state of near polar alignment, with the inclination somewhat less than $90^\circ$.}
\item The timescale of evolution of the disk-binary inclination angle [see Eqs.~\eqref{eq:dcosdbdt}-\eqref{eq:dcosdedt}] depends on various disk parameters [see Eq.~\eqref{eq:cgb_scale}], but is in general less than a few Myrs. This suggests that highly inclined disks and planets may exist around eccentric binaries.
\end{enumerate}

\section*{Acknowledgments}

% J.J.:  Dong, please do not take out my thank you's to Lubow, Martin, and Li.

JZ thanks Stephen Lubow, Rebbeca Martin, and Gongjie Li for useful conversations. { We thank the anonymous referee for their comments which improved the quality of this paper.  }  This work has been supported in part by NASA grants NNX14AG94G and
NNX14AP31G, and a Simons Fellowship from the Simons Foundation.  JZ is
supported by a NASA Earth and Space Sciences Fellowship in
Astrophysics.

\section*{Appendix: Accretion Torques}

\begin{figure*}
\includegraphics[scale=1.0]{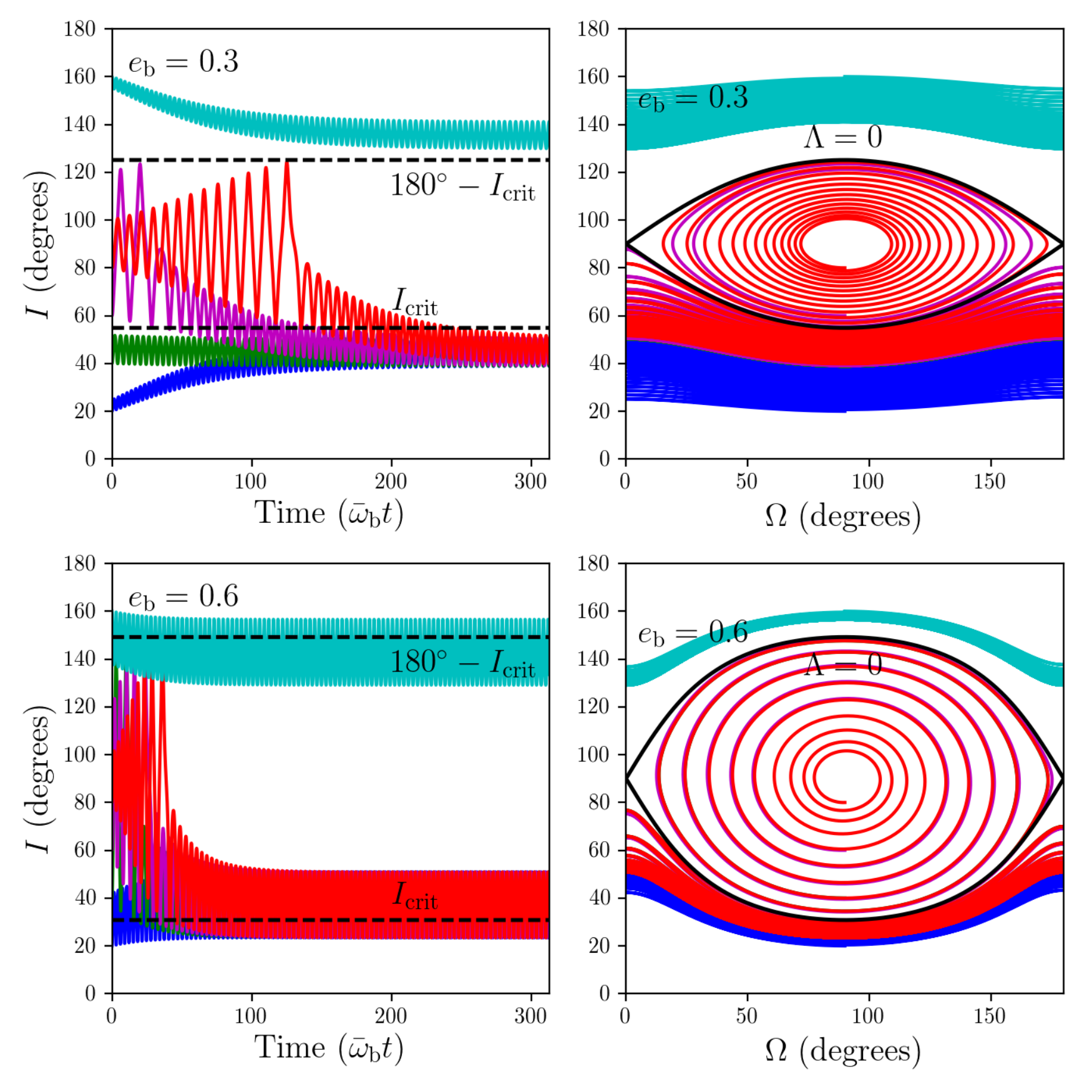}
\caption{Same as Fig.~\ref{fig:visc}, except we neglect the disk's viscous torque [Eq.~\eqref{eq:dldt_visc}], and include the disk's accretion torque [Eq.~\eqref{eq:dldt_acc}].  All parameter values are the same, except $\ag = 0.2$, $h = 0.3$, and $\lam = 1$.}
\label{fig:acc}
\end{figure*}

If the inner disk is not coplanar with the outer disk, accretion will change the disk angular momentum vector over time.  We parameterize this accretion torque according to
\be
\left( \frac{\der \bLd}{\der t} \right)_{\rm acc} = -\lam \dot M \sqrt{ G \Mb \rin} \bl(\rin,t),
\label{eq:Tacc}
\ee
where $\dot M$ is the accretion rate onto the binary, and $\lam \sim 1$ parameterizes the angular momentum loss from the disk to the binary.  The time evolution of $\bG$, as well as the pericenter precession induced by the non-Keplarian angular frequency, warps the inner edge of the disk in the direction of the binary orbital plane [$(\blone)_{\rm warp}$; Eq.~\eqref{eq:lone_warp}].  Inserting $\bl(r,t) = \ld + (\blone)_{\rm warp}$ in equation~\eqref{eq:Tacc}, we obtain
\begin{align}
\bigg( \frac{\der \ld}{\der t} \bigg)_{\rm acc} =  \cga \bomb \tb(\rin) \Lam &\Big[ (1-\eb^2) (\ld \bcdot \blb) \ld \btimes (\blb \btimes \ld)
\nonumber \\
&- 5 (\ld \bcdot \beb) \ld \btimes (\beb \btimes \ld) \Big]
\nonumber \\
+ \cga \Wbb(\rin) \fb & \Big[ (1-\eb^2) (\ld \bcdot \blb) \ld \btimes (\blb \btimes \ld)
\nonumber \\
&- 5(\ld \bcdot \beb) \ld \btimes (\beb \btimes \ld) \Big],
\label{eq:dldt_acc}
\end{align}
where
\begin{align}
\cga &= \frac{\lam \dot M \sqrt{G \Mb \rin}}{\Ld} \simeq \frac{9}{4} \lam \ag h^2 n(\rout) 
\nonumber \\
=& \ 3.18 \times 10^{-7} \lam \left( \frac{\ag}{0.01} \right) \left( \frac{h}{0.1} \right)^2
\nonumber \\
&\times \left( \frac{\Mb}{2 \, M_\odot} \right)^{1/2} \left( \frac{100 \, \text{AU}}{\rout} \right)^{3/2} \left( \frac{2\pi}{\text{yr}} \right),
\label{eq:cga}
\end{align}
We have assumed the disk to be in a steady state, so
\be
 \dot M \simeq 3\pi \ag h^2\Sgin \sqrt{G \Mb \rin}.
 \label{eq:dotM}
 \ee
 Equation~\eqref{eq:dldt_acc} agrees with the rough magnitude and direction of the accretion torque estimated in Equation~\eqref{eq:dldt_acc_mag}.
 
Since
\begin{align}
\left.\frac{\der}{\der t} (\ld \bcdot \blb) \right|_{\rm acc} = \ &\cga \bomb \tb(\rin) \Lam \big[ (1-\eb^2) - \Lam \big]
\nonumber \\
&+ \cga \Wbb(\rin) \fb \big[ (1-\eb^2) - \Lam \big]
\label{eq:acc_1} \\
\left. \frac{\der}{\der t}(\ld \bcdot \beb) \right|_{\rm acc} = & -\cga \bomb \tb(\rin) \Lam \big[ \Lam + 5 \eb^2 \big] 
\nonumber \\
&- \cga \Wbb(\rin) \fb \big[ \Lam + 5 \eb^2 \big],
\label{eq:acc_2}
\end{align}
the radial functions $\tb(\rin), \Wbb(\rin) < 0$, and $\fb \sim \Lam$, Eqs.~\eqref{eq:acc_1}-\eqref{eq:acc_2} drives the disk one of two ways depending on the rough value of $\Lam$:
\begin{enumerate}
\item $\Lam \gtrsim 0$: The accretion torque~\eqref{eq:dldt_acc} pushes $\ld$ away from~$\blb$.
\item $\Lam \lesssim 0$: The accretion torque~\eqref{eq:dldt_acc} pushes $\ld$ away from~$\beb$.
\end{enumerate}
From Eq.~\eqref{eq:dl0dt}, it may be shown that there are no fixed points near the $\Lam = 0$ separatrix.  Therefore, the accretion torque drives the disk to a trajectory near the $\Lam = 0$ separatrix. 

Figure~\ref{fig:acc} plots the examples considered in Figure~\ref{fig:cons} with accretion torques [Eq.~\eqref{eq:dldt_acc}].  We take $h$ and $\ag$ to be significantly higher than our cannonical values of $\ag = 0.01$ and $h = 0.1$ so that accretion torques effect the dynamical evolution of the circumbinary disk [Eq.~\eqref{eq:cga}].  In the left panels of Figure~\ref{fig:acc}, we plot the disk inclination with time, for the binary eccentricities indicated.  The trajectories which start at $I(0) = 20^\circ, 40^\circ,$ and $80^\circ$ all evolve toward the prograde seperatrix, which nutates around $I \sim 50^\circ$ when $\eb = 0.3$, and $I \sim 40^\circ$ when $\eb = 0.6$.  The trajectories which start at $I(0) = 60^\circ$ both evolve to the retrograde seperatrix, which nutates around $I \sim 130^\circ$ when $\eb = 0.3$, and $I \sim 140^\circ$ when $\eb = 0.6$.  On the right panels, we plot the disk trajectories on the $I-\Om$ plane, for the binary eccentricities indicated.  All disk trajectories evolve toward the $\Lam \approx 0$ seperatrix.

\begin{figure*}
\includegraphics[scale=1.0]{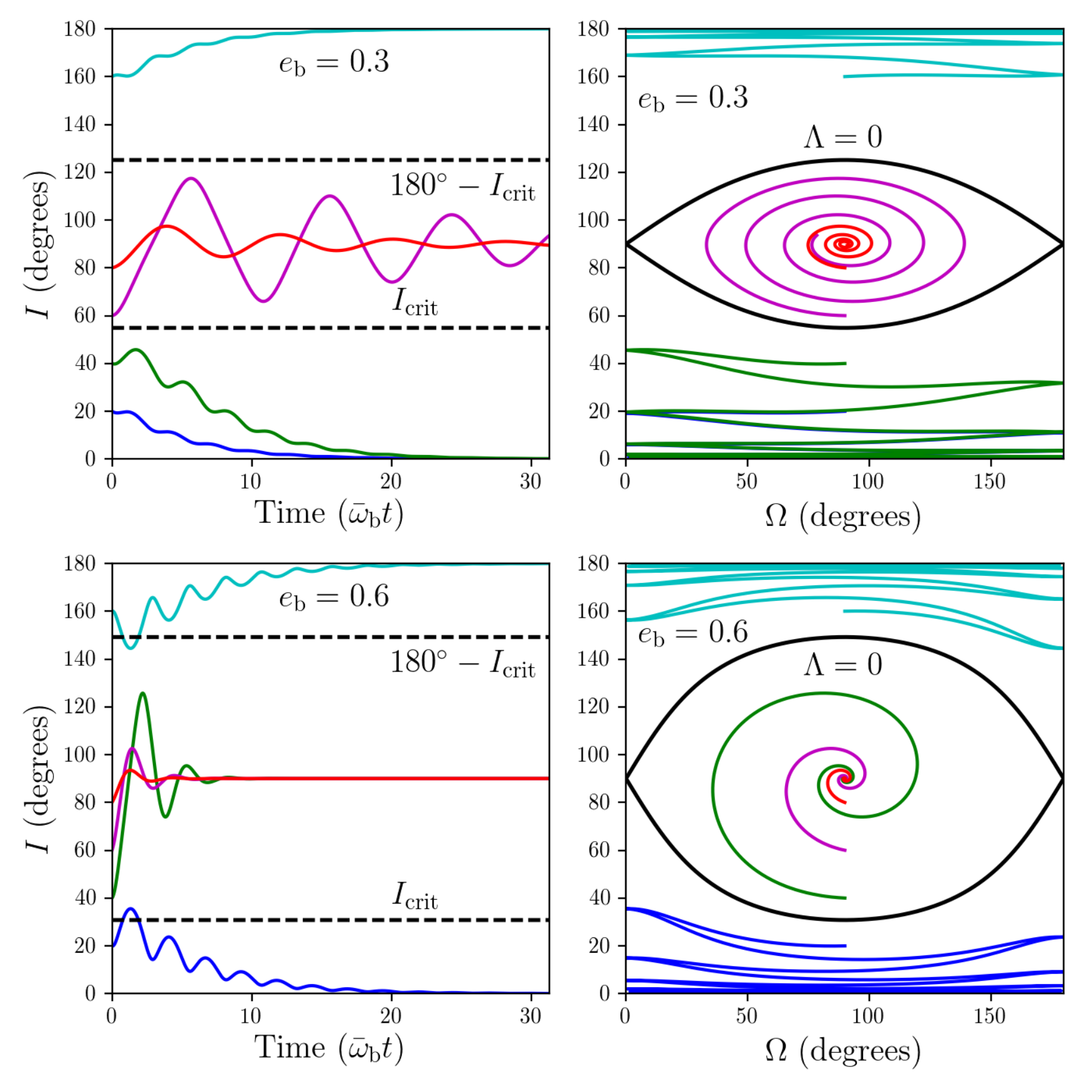}
\caption{Same as Figure~\ref{fig:visc}, except we include the disk's accretion torque [Eq.~\eqref{eq:dldt_acc}].  All parameter values are the same, with $\lam = 1$.}
\label{fig:both}
\end{figure*}

The relative strength of the viscous to the accretion torques from disk warping is given by the ratio
\be
\frac{|\cgb|}{|\cga \Wbb(\rin)|}\approx 300 \lam^{-1} \left( \frac{0.1}{h} \right)^2.
\ee
As long as $|\cgb| \gg |\cga \Wbb(\rin)|$, the viscous torque dominates, and $\ld$ aligns with either $\blb$ or $\beb$, depending on the sign of $\Lam$ (Sec.~\ref{sec:Visc}).  When $|\cgb| \lesssim |\cga \Wbb(\rin)|$, the accretion torques may dominate, and $\ld$ may be driven to the seperatrix $\Lam \approx 0$.

Figure~\ref{fig:both} is identical to Figure~\ref{fig:visc}, except we include viscous [Eq.~\eqref{eq:dldt_visc}] and accretion [Eq.~\eqref{eq:dldt_acc}] torques with $\ag = 0.01$, $h = 0.1$, and $\lam = 1$.  Because $|\cgb| \gg |\Wbb(\rin)\cga|$, the viscous torque dominates the disk's dynamics.  As a result, Figure~\ref{fig:both} is almost indistinguishable from Figure~\ref{fig:visc}.  Only for unrealistically hot protoplanetary disks with $h\gtrsim 0.5$ may accretion torques significantly effect the disk evolution over viscous timescales.

\end{document}